\documentclass[letterpaper,prx,aps,floatfix,notilepage,nofootinbib, superscriptaddress, reprint]{revtex4-2}

\usepackage[english]{babel}
\usepackage{amsthm}
\usepackage{amssymb}
\usepackage{mathtools}

\usepackage{etoolbox}
\usepackage{bbold}

\usepackage{comment}
\def\bs#1{{\textbf{#1}}}
\def\tr{{\textrm{tr}}}
\def\({\left(}
\def\){\right)}

\usepackage{graphicx}
\usepackage{dcolumn}
\usepackage{bm}
\usepackage{appendix}

\usepackage{xr}
\makeatletter
\newcommand*{\addFileDependency}[1]{
\typeout{(#1)}
\@addtofilelist{#1}
\IfFileExists{#1}{}{\typeout{No file #1.}}
}\makeatother

\newcommand*{\myexternaldocument}[1]{%
\externaldocument{#1}%
\addFileDependency{#1.tex}%
\addFileDependency{#1.aux}%
}
\myexternaldocument{SI}

\usepackage[
    colorlinks=true,
    citecolor=blue,
    linkcolor=blue,
    urlcolor=blue]{hyperref}

\usepackage{xcolor}

\newcommand{\ket}[1]{|#1\rangle}
\newcommand{\bra}[1]{\langle#1|}

\newcommand{\expect}[1]{\mathbb{E}#1}

\newtheorem{theorem}{Theorem}[section]
\newtheorem{lemma}[theorem]{Lemma}

\begin{document}

\title{Experimental demonstration of scalable cross-entropy benchmarking to detect measurement-induced phase transitions on a superconducting quantum processor}

\author{Hirsh Kamakari}
\affiliation{Division of Engineering and Applied Science, California Institute of Technology, Pasadena, CA 91125, USA}

\author{Jiace Sun}
\affiliation{Division of Engineering and Applied Science, California Institute of Technology, Pasadena, CA 91125, USA}

\author{Yaodong Li}
\email{liyd@stanford.edu}
\affiliation{Department of Physics, Stanford University, Stanford, CA 94305, USA}

\author{Jonathan J. Thio}
\affiliation{Department of Physics, University of Cambridge, Cambridge, CB3 0HE, UK}

\author{Tanvi P. Gujarati}
\affiliation{IBM Quantum, IBM Research Almaden, San Jose, CA 95120, USA}

\author{Matthew P. A. Fisher}
\affiliation{Department of Physics, University of California, Santa Barbara, CA 93106, USA}

\author{Mario Motta}
\affiliation{IBM Quantum, IBM Research Almaden, San Jose, CA 95120, USA}
\affiliation{IBM Quantum, IBM T. J. Watson Research Center, Yorktown Heights, NY 10598, USA}

\author{Austin J. Minnich}
\email{aminnich@caltech.edu}
\affiliation{Division of Engineering and Applied Science, California Institute of Technology, Pasadena, CA 91125, USA}

\date{March 25, 2025}

\begin{abstract}
Quantum systems subject to random unitary evolution and measurements at random points in spacetime exhibit entanglement phase transitions which depend on the frequency of these measurements. Past work has experimentally observed entanglement phase transitions on near-term quantum computers, but the characterization approach using entanglement entropy is not scalable due to
exponential  
overhead
of quantum state tomography and post-selection.
Recently, an alternative protocol to detect entanglement phase transitions using linear cross-entropy was proposed, attempting to eliminate both bottlenecks.
Here, we report demonstrations of this protocol in systems with one-dimensional and all-to-all connectivities
on IBM's quantum hardware on up to 22 qubits, a regime which is presently inaccessible if post-selection is required.
We demonstrate data collapses onto scaling functions with critical exponents {in semi-quantitative agreement with theory.} Our demonstration  of the cross entropy benchmark (XEB) protocol paves the way for studies of measurement-induced entanglement phase transitions and associated critical phenomena on larger near-term quantum systems.

\end{abstract}

\maketitle


Quantum systems undergoing unitary evolution in the presence of an observer making measurements (monitored quantum systems)~\cite{nahum2018hybrid, nandkishore2018hybrid, li2018hybrid} exhibit unique dynamics, distinct from both thermalizing closed systems~\cite{Nandkishore2015} and conventional open quantum systems~\cite{Breuer2002theory}.
When the system is weakly monitored and subject to sufficiently entangling unitaries, initial product states typically exhibit a linear in time growth of the entanglement entropy, before evolving into steady states where the entanglement entropy admits a volume-law scaling~\cite{Calabrese_2005, Calabrese_2007, Kim2013, nahum2018operator, keyserlingk2018operator}. 
In contrast, strongly monitored systems 
are not able to support highly entangled states, resulting in area-law entanglement scaling at long times~\cite{misra1977, cao2018monitoring}.
Separating the two phases lies a phase transition, which was initially found theoretically in simplified quantum circuit models with mid-circuit measurements, and was later found to be generic to a wide range of monitored dynamics ~\cite{cao2018monitoring, li2019hybrid, nahum2019majorana,  vasseur2020mft, barkeshli2020symmetric,  sang2020protected,ippoliti2020measurementonly, chenxiao2020nonunitary,ashida2020continuous,diehl2020trajectory, sagar2020volume,  nahum2020alltoall,   bao2021enriched, vasseur2021chargesharpening, barratt2021sharpening, agrawal2023observing}.
Such measurement-induced phase transitions (MIPTs) have recently garnered much interest~\cite{choi2019qec, choi2019spin, gullans2019purification, gullans2020lowdepth, harrow2020efficient, Fisher_2023_annrevcmp}.

An experimental observation of MIPTs was recently demonstrated on IBM quantum hardware with up to 14 qubits~\cite{koh2022}. By directly measuring the entanglement entropy
after a comprehensive quantum tomography of the steady states, Koh et.~al.~\cite{koh2022} were able to observe an MIPT and confirm the competing effects of random unitaries and mid-circuit measurements. However, the
experiment required over 5200 device-hours and is limited in scalability due to the exponential cost of quantum state tomography  and post-selection of measurement outcomes.
The lifetime of superconducting qubits also puts a stringent limit on the circuit depth (as well as on system size when circuit depth scales with the number of qubits), since mid-circuit measurements can be several times slower than two-qubit unitary gates.

To avoid mid-circuit measurements, a space-time duality mapping was introduced~\cite{ippoliti2021postselection, lu2021spacetime} and recently implemented on Google's superconducting processor~\cite{google2023measurement},  
where phase transitions were observed in 1D unitary circuits with a reduced number of post-selections, and at the boundary of shallow 2D unitary circuits of $70$ qubits without post-selection.
Alternatively, order parameters based on reference qubits can be used to efficiently and scalably probe MIPTs~\cite{gullans2020scalable}, where post-selection can be avoided with an accompanying classical simulation.
The use of a reference qubit to probe MIPTs has been demonstrated in trapped ion systems for Clifford circuits~\cite{monroe2021TrappedIonCliffordTransition}, featuring a high gate fidelity and non-local qubit connectivity. 
Another order parameter which can be used to probe MIPTs is the cross entropy \cite{li2023}, which requires no ancilla qubits or the exponential overhead of post-selection.
Although the theoretical basis for this method has been established, a demonstration of this protocol on near-term quantum hardware has not yet been reported.

In this Letter, we report an experimental demonstration of the detection of MIPTs on hybrid Clifford circuit models with up to 22 physical qubits.
The required circuits were executed in less than 8 device-hours on IBM superconducting devices, representing a decrease in device time by nearly two orders of magnitude compared to the approach based on measuring entanglement entropy. 
Moreover, a circuit compression technique allows us to investigate circuit models with all-to-all connectivity on IBM's 2D layout.
From the data, we extract critical exponents which are in {semi-quantitative agreement with theoretical predictions.}
This work paves the way for studies of other quantum critical phenomena on near-term quantum hardware using the XEB protocol and provides a potential benchmarking tool for quantum circuits with mid-circuit measurements.

\begin{figure}
    \centering
    \includegraphics[width=.9\linewidth]{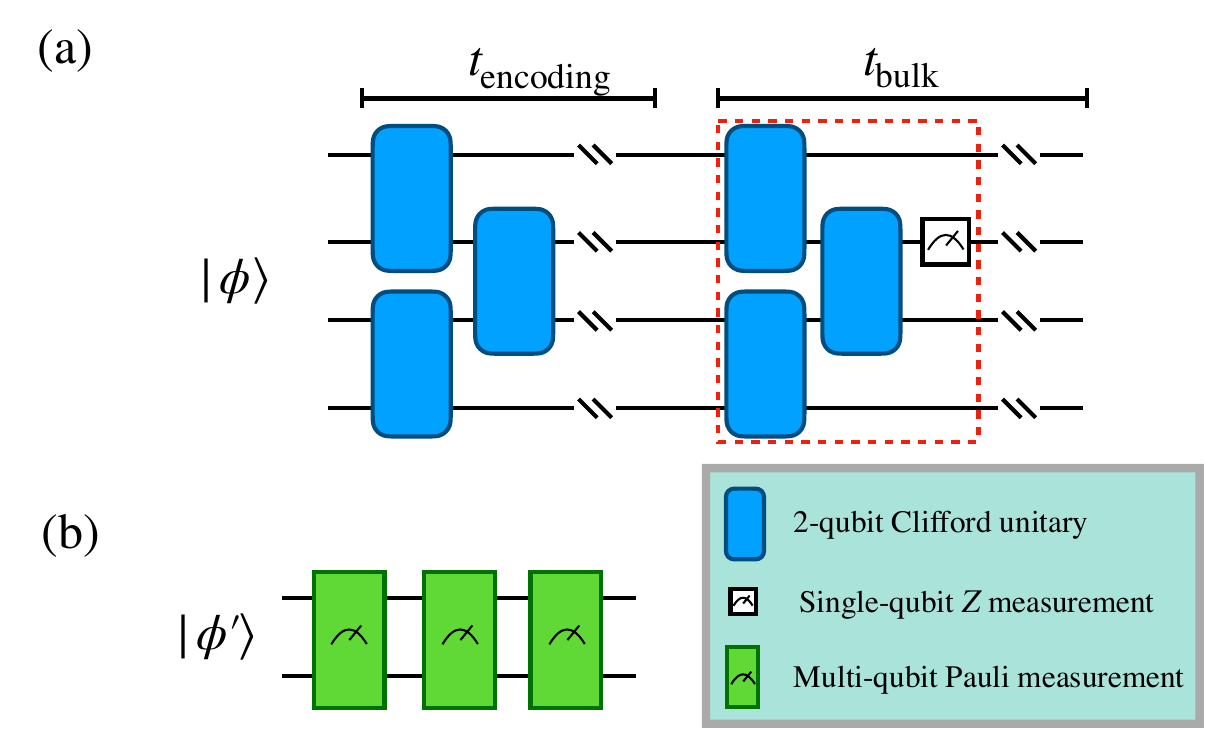}
    \caption{
    Schematic of the protocol demonstrated in this Letter. 
    (a) An $L$-qubit Clifford circuit used in the cross entropy benchmark protocol \cite{li2023}.
    We choose the initial state $\ket{\phi}$ to be either $\ket{0T}^{\otimes L/2}$, where $\ket{T}$ is a magic state, or $\ket{0}^{\otimes L}$. The red box represents one layer of unitaries (and one layer of measurements in the bulk stage).
    (b) The compressed $L/2$-qubit circuit consisting of at most $L/2$ multi-qubit Pauli measurements. The compressed initial state is $\ket{\phi'}=\ket{T}^{\otimes L/2}$ or $\ket{0}^{\otimes L/2}$.
    }
    \label{fig:circuit}
\end{figure}

\textit{Circuit model and cross entropy benchmark ---} We begin by describing the cross entropy benchmark protocol of Ref.~\cite{li2023}.
We consider a family of random circuits, where each circuit consists of two stages: an purely unitary ``encoding stage'' consisting of $t_{\mathrm{encoding}}$ layers, and a ``bulk stage'' consisting of $t_{\mathrm{bulk}}$ layers with both unitary gates and mid-circuit measurements, see Fig.~\ref{fig:circuit}(a).
For an $L$-qubit circuit, both stages must contain a number of layers scaling at least linearly with $L$ for the system to enter a steady state, particularly when the steady state has volume-law scaling of entanglement entropy.

The protocol involves the application of the same circuit to two different initial states, $\rho$ and $\sigma$, and a comparison between the two ensembles of measurement records.
For a given circuit $C$ with $N$ mid-circuit measurements, a measurement record $\mathbf{m}=(m_1,m_2,\dots,m_N)$, where $m_j$ are the outcomes (0 or 1) from each mid-circuit measurement, is sampled by running the circuit on quantum hardware with input state $\rho$.
The sampled measurement records obey a probability distribution, which we denote as $p_\mathbf{m}^\rho$.
We also implement the sampling experiment using classical simulations of the same circuit, but with a different (stabilizer) initial state $\sigma$.
The corresponding measurement record probabilities is similarly denoted as $p_\mathbf{m}^\sigma$.
The normalized linear cross entropy acts as a distance measure between the two distributions, which we define for the circuit $C$ as 
\begin{equation}
    \label{eq:def_linear_XEB}
     \chi_C=\frac{\sum_\mathbf{m}p_\mathbf{m}^\rho p_\mathbf{m}^\sigma}{\sum_\mathbf{m}(p_\mathbf{m}^\sigma)^2},
\end{equation}
which can be estimated by taking the sample average of $p_\mathbf{m}^\sigma / \left( \sum_\mathbf{m}(p_\mathbf{m}^\sigma)^2 \right)$ 
over many runs of the quantum circuit with input state $\rho$.
We then average over random circuits $C$, obtaining the final cross entropy for a given measurement rate as $\chi=\mathbb{E}_C\chi_C$.

As shown in~\cite{li2023}, for $\rho \neq \sigma$ and in the absence of noise, the quantity $\chi$ acts as an order parameter which, in the thermodynamic limit, approaches 1 when the system is in the volume-law phase and approaches a constant strictly less than 1 in the area-law phase.
Intuitively, $\chi$ measures the distinguishability of the two initial states by comparing mid-circuit measurement records, after the two initial states are ``scrambled'' by the encoding unitary.
The linear cross entropy has been used as a figure of merit for random circuit sampling~\cite{boixo2016XEB, google2019supremacy, dalzell2021random, gao2021limitations}, and has since become a standard tool for benchmarking quantum simulators~\cite{choi2023preparing, mark2023benchmarking}.



For $\chi$ to be efficiently obtainable from this protocol, the circuit for which sampling is performed classically needs to be efficiently classically computable.  This is possible when the circuit contains only Clifford operations and when the input state $\sigma$ is a stabilizer state. However, if the $\rho$ is not a stabilizer state, then the overall protocol cannot be efficiently simulated classically in the asymptotic limit of system size.  The cross entropy protocol is similar in spirit to hybrid quantum-classical observables used in previous experiments~\cite{google2023measurement, monroe2021TrappedIonCliffordTransition} (see also~\cite{garratt2022measurements, feng2022measurementinduced}) and, as we will show, allows us to probe the transition and obtain critical exponents on noisy processors.

\textit{Experimental implementation ---} We implemented this approach on IBM Quantum processors.
The systems we consider are a 1D chain with nearest-neighbor qubit connectivity and an infinite-dimensional system with all-to-all qubit connectivity. We chose the initial $L$-qubit states on the quantum processor in both cases to be $\rho=|0T0T\cdots 0T\rangle\langle T0T0\cdots T0|$ with $|T\rangle = (|0\rangle + \exp{(i\pi/4)}|1\rangle)/\sqrt{2}$, the alternating magic state, and $\sigma=|0^{\otimes L}\rangle\langle 0^{\otimes L}|$, the all-zero state.
Note that $\rho$ is not a stabilizer state. For the alternating magic states, the number of $T$ gates grows linearly with the number of qubits, so that an exact simulation of the circuit via either the state-vector simulation or Clifford+T simulation~\cite{josza_vandennest, koh2015extensions, bouland2018conjugated, bravyi2016improved} requires exponential classical resources in the asymptotic limit of large $L$.

For all experiments, the circuits are constructed using alternating layers of unitaries and measurements. Each unitary layer consists of $L/2$ two-qubit unitary gates, sampled uniformly from the two-qubit Clifford group. 
For the 1D chain, the two-qubit Clifford gates are applied in a brickwork pattern on nearest-neighbor qubits. For the infinite dimensional system, $L/2$ two-qubit unitaries are applied to pairs of qubits selected uniformly at random. Each measurement layer, in both the 1D chain and the infinite-dimensional system, consists of single-qubit $Z$ measurements occurring on each qubit with probability $p$. 
For both systems, we used an encoding ratio and bulk ratio of 3, namely $t_{\mathrm{bulk}}=t_{\mathrm{encoding}}=3L$.

The resulting circuits with the above properties have as many as $L^2$ mid-circuit measurements, which are relatively slow operations and introduce both readout and quantum state errors, and so they cannot be executed while preserving adequate fidelity on the current platforms.
We therefore employ a circuit compression scheme which exploits the input state being an alternating magic state and the circuit bulk being fully Clifford~\cite{SM, yoganathan2019}.
Such a circuit compression scheme converts the input Clifford circuit into the Pauli-based computing model (PBC) which is composed of only Pauli measurements~\cite{PhysRevX.6.021043}.
The PBC model can then be recast as a standard circuit with a reduced number of gates and measurements on a subset of the qubits.
After circuit compression, we obtain a circuit with $L/2$ hardware qubits and at most $L/2$ multi-qubit Pauli measurements, significantly fewer than an average of $pt_{\mathrm{bulk}}L^2$ measurements in the original uncompressed circuits.
The initial state of the circuit is now $\ket{T}^{\otimes L/2}$, see Fig.~\ref{fig:circuit}(b).
All circuits used in our experiments use Clifford compression, allowing us to treat up to 44 qubit systems using only 22 physical qubits. We emphasize, however, that the vast majority of the computational cost reduction compared to Ref.~\cite{koh2022} arises from the XEB protocol owing to the elimination of post-selection and tomography, both of which incur exponential overheads.

For the 1D-chain experiments, qubits were selected heuristically at submission time based on the one-qubit gate, two-qubit gate, and readout error rates provided by IBM in their hardware calibration data.
We minimize the average errors that would occur in all circuits based on the number and placement of gates and measurements in the circuits~\cite{SM}.
For the all-to-all experiments, we used the same qubit layouts that were selected for the 1D-chain. For all experiments, we generated 1000 random circuits for each $(L,p)$ pair, and each circuit was run 1000 times, all on the \textit{ibm\_sherbrooke} machine.

\begin{figure}
    \centering
    \includegraphics[width=\linewidth]{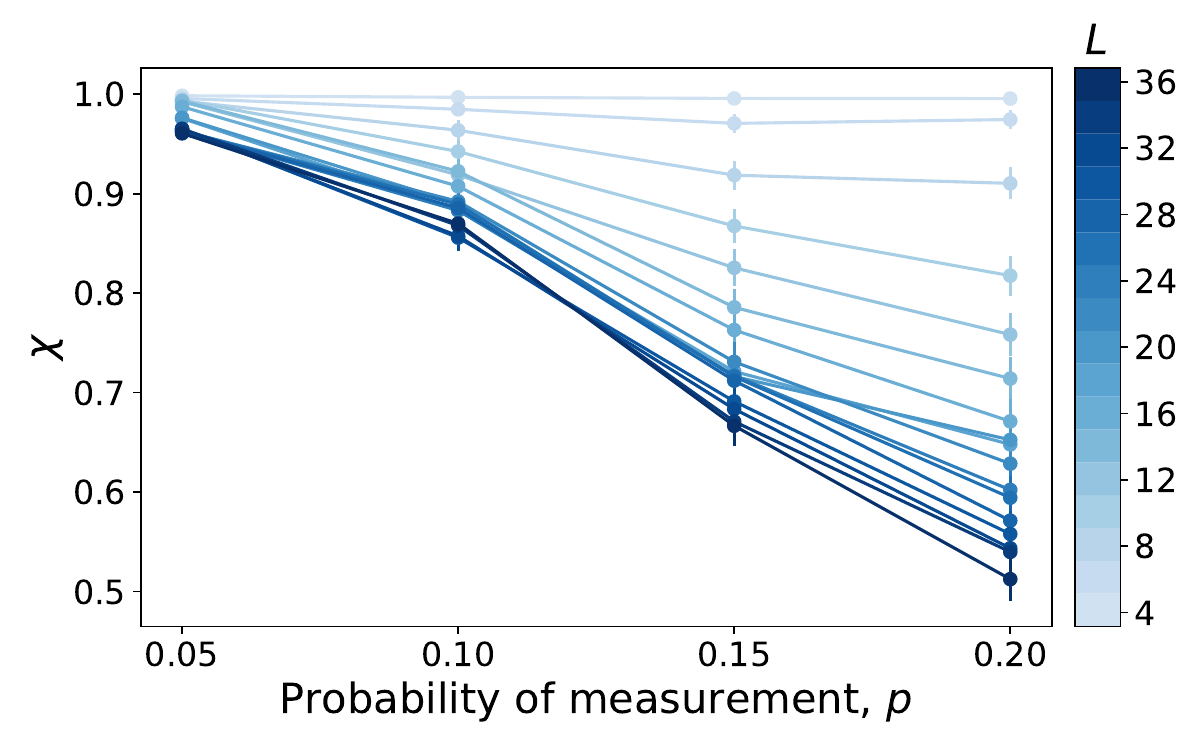}
    \caption{Cross entropy for identical initial states ($\rho=\sigma)$ obtained from \textit{ibm\_sherbrooke} with up to 18 physical qubits (equivalent to a system size of $L=36$ qubits before compression). 
    The errors incurred from the physical qubits results in a cross entropy lower than the theoretical value of 1. 
    }
    \label{fig:rho=sigma-cross-entropy}
\end{figure}

\begin{figure}[htb!]
        \includegraphics[width=0.45\textwidth]{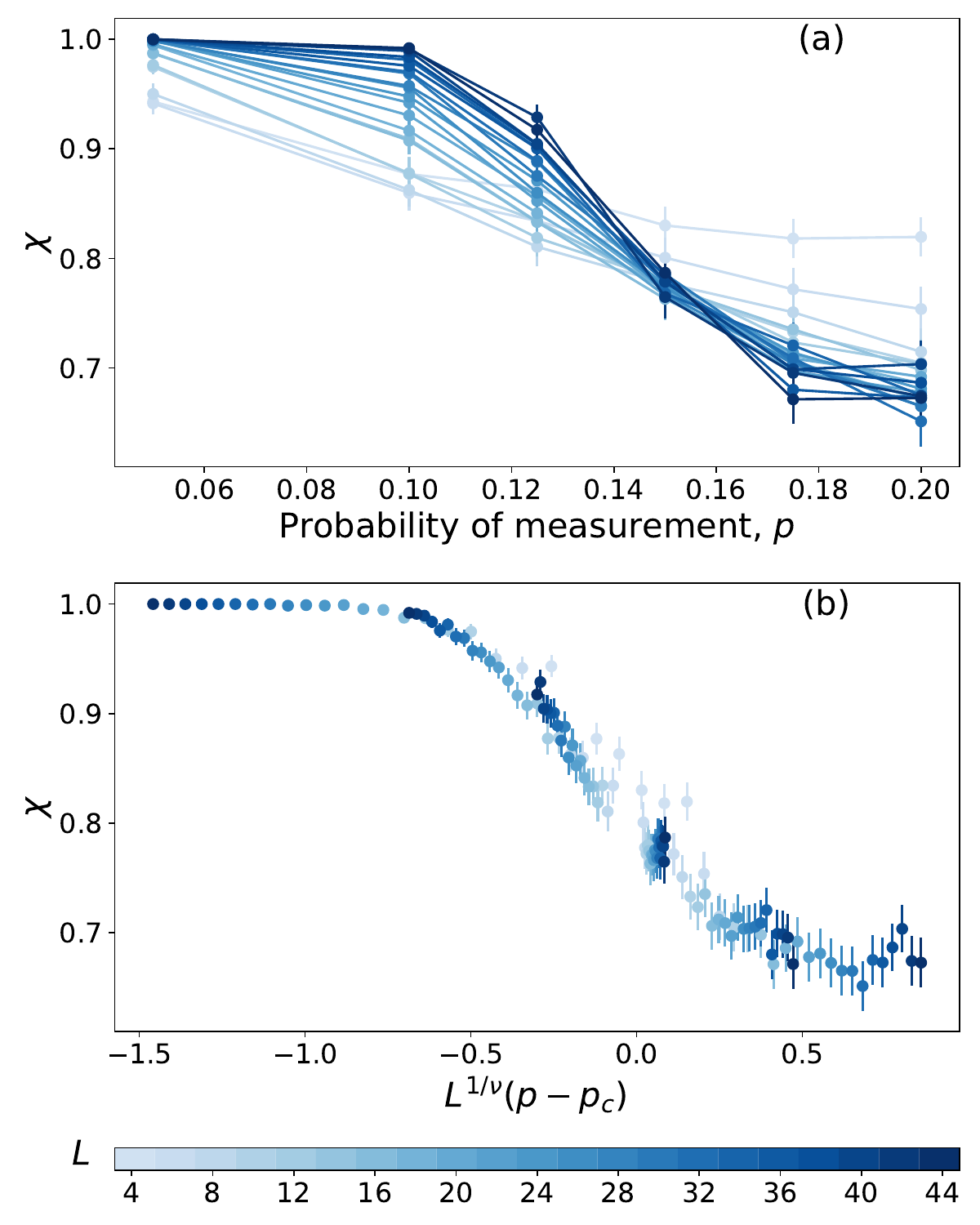}
\caption{(a) Cross entropy $\chi$ for 1D chains with up to 22 physical qubits (corresponding to a system size of $L=44$ qubits before compression) computed on \textit{ibm\_sherbrooke}. (b) Collapse of cross entropy curves near the critical point obtained by minimizing the scatter of all points to an unknown scaling function. The fitting procedure gives a critical measurement rate of $p_c=0.14\pm0.01$ and critical exponent $\nu=1.4 \pm 0.5$.}
\label{fig:1D-chain-cross-entropy}
\end{figure}

\textit{Results for 1D connectivity} ---
We first present the experimental results when we set $\rho=\sigma$ to provide a benchmark of the hardware performance. 
We obtain the circuits from the compressed 1D circuits, but replace all $\ket{T}$ states with $\ket{0}$ states, so that the initial states on both $\rho$ and $\sigma$ are the all-zero state, see Fig.~\ref{fig:circuit}(b).
In this case, since the circuits run on both the quantum and classical sides are identical, we expect to observe $\chi=1$ for all $L$ and for all $p$ in the absence of any noise or hardware errors. The deviation of the cross entropy from unity therefore provides a measure of the overall errors and noise in the circuit, which could be due to various sources such as gate errors, qubit decay and dephasing, and cross-talk from mid-circuit measurements. 

Fig.~\ref{fig:rho=sigma-cross-entropy} shows the measured cross entropy $\chi$ versus $p$ for various $L$.  We observe that $\chi > 0.9$ for $L\leq 8$ for all $p$, and due to hardware noise $\chi$ decreases to well below unity for larger $L$ and $p \gtrsim 0.1$.
That $\chi \approx 1$ for small $L \leq 8$ can be attributed to the short circuit depths of these circuits. For instance, a compressed four qubit circuit requires only two physical qubits and two Pauli measurements, allowing them to be executed with high fidelity and resulting in $\chi > 0.9$.
For increasing $L$ at fixed $p$, both uncompressed and compressed circuits contain more mid-circuit measurements.  

From Fig.~\ref{fig:rho=sigma-cross-entropy}, we observe an overall trend that $\chi$ decreases when either $L$ or $p$ is increased.
This observation is in qualitative agreement with our numerical simulations of a noisy circuit model with a depolarizing channel applied at each spacetime location with probability $q=0.1$\%~\cite{SM}. The qualitative agreement of the experimental and simulated $\chi$ for the $\rho=\sigma$ case suggests that the primary characteristics of the noise are captured by a simple depolarizing channel. We note that the value of $q=0.1$\% is of the order of values  inferred for superconducting quantum computers in prior works, including those from IBM~\cite{koh2022, PRXQuantum.3.040318}.

We next present experimental results for the 1D chain for $\rho \neq \sigma$.
Fig.~\ref{fig:1D-chain-cross-entropy} shows $\chi$ versus $p$ for $L$ between 4 and 44 (corresponding to 2 to 22 physical qubits)  up to $p=0.2$
obtained from the 127 qubit $ibm\_sherbooke$ device.
Qualitatively, we see the expected characteristics  as described below Eq.~\eqref{eq:def_linear_XEB}; namely, with increasing $L$, $\chi$ approaches unity for $p \lesssim 0.12$, while it plateaus to a constant $<1$ for larger values of $p$.
The curves for different $L$ cross at a value of $p$ we denote as $p_c$, with $p_c \in (0.15, 0.175)$. The crossing indicates the occurrence of a MIPT as theoretically discussed in Ref.~\cite{li2023}. We note that the crossing is also observed in the noisy classical simulations in the Supplemental Material~\cite{SM}, suggesting that the MIPT is robust to noise at the $q=0.1$\% scale.

The cross entropy $\chi$ is related to a domain wall free energy in an associated statistical mechanics model~\cite{li2023}, and its value near the critical point depends only on the ratio of the system size and the correlation length, according to standard scaling hypotheses. We verify this hypothesis by collapsing the data from different system sizes $L$ and measurement probabilities $p$ to an unknown but universal scaling function $F$:
\begin{equation}
    \chi(L, p)=F\left[L^{1/\nu}(p-p_c)\right],
    \label{eq:finite-size-scaling}
\end{equation}
where $\nu$ is the critical exponent that controls the divergence of the correlation length, and $p_c$ is the critical measurement rate~\cite{stanley1999}.
With $F$ unknown, we follow standard methods~\cite{koh2022, bhattacharjee2001} to choose the parameters $p_c$ and $\nu$ so as to optimize the quality of the data collapse~\cite{SM}. 

The resulting collapsed curve is shown in Fig.~\ref{fig:1D-chain-cross-entropy}(b). We obtain the critical measurement rate $p_c=0.14\pm0.01$ and critical exponent $\nu=1.4 \pm 0.5$ at the 90\% confidence level.
As in the Supplemental Material~\cite{SM}, our reported values of $p_c$ and $\nu$ are in quantitative agreement with
classical numerical calculations of uncompressed circuits in the presence of $0.1\%$ erasure noise, where we obtained $\nu\approx1.33$ and $p_c=0.14$; we also find that noise does not qualitatively change the critical parameters from their noiseless values.
For chains of fewer than 10 qubits, finite-size effect are observed as indicated by deviations from the collapsed curve as well as the plateau to a larger value for large $p$. Removing the smaller system sizes from the fitting did not change the values of $\nu$ and $p_c$ within the reported uncertainties.

\textit{Results for all-to-all connectivity} ---
We finally present the experimental results for the all-to-all connectivity experiment. Compared to 1D systems, all-to-all connected systems without compression would require $O(L^3)$ SWAP gates per circuit to implement all the 2 qubit unitaries on hardware with nearest neighbor interactions. This prohibitive scaling makes all-to-all systems harder to simulate than 1D systems. With circuit compression, however, the resource requirements are the same as the 1D system since in both cases the compressed circuits have $L/2$ qubits and at most $L/2$ mid-circuit measurements. We demonstrate this resource reduction application of circuit compression by experimentally observing an MIPT for an all-to-all connected system. Theory predicts qualitatively similar dependencies of the cross entropy on $p$ and $L$ as in the 1D case, but the transition is in a different universality class~\cite{nahum2021}. The initial states used in this experiment are the same as in the 1D-chain case. 

\begin{figure}[htb!]
    \centering
        \includegraphics[width=0.45\textwidth]{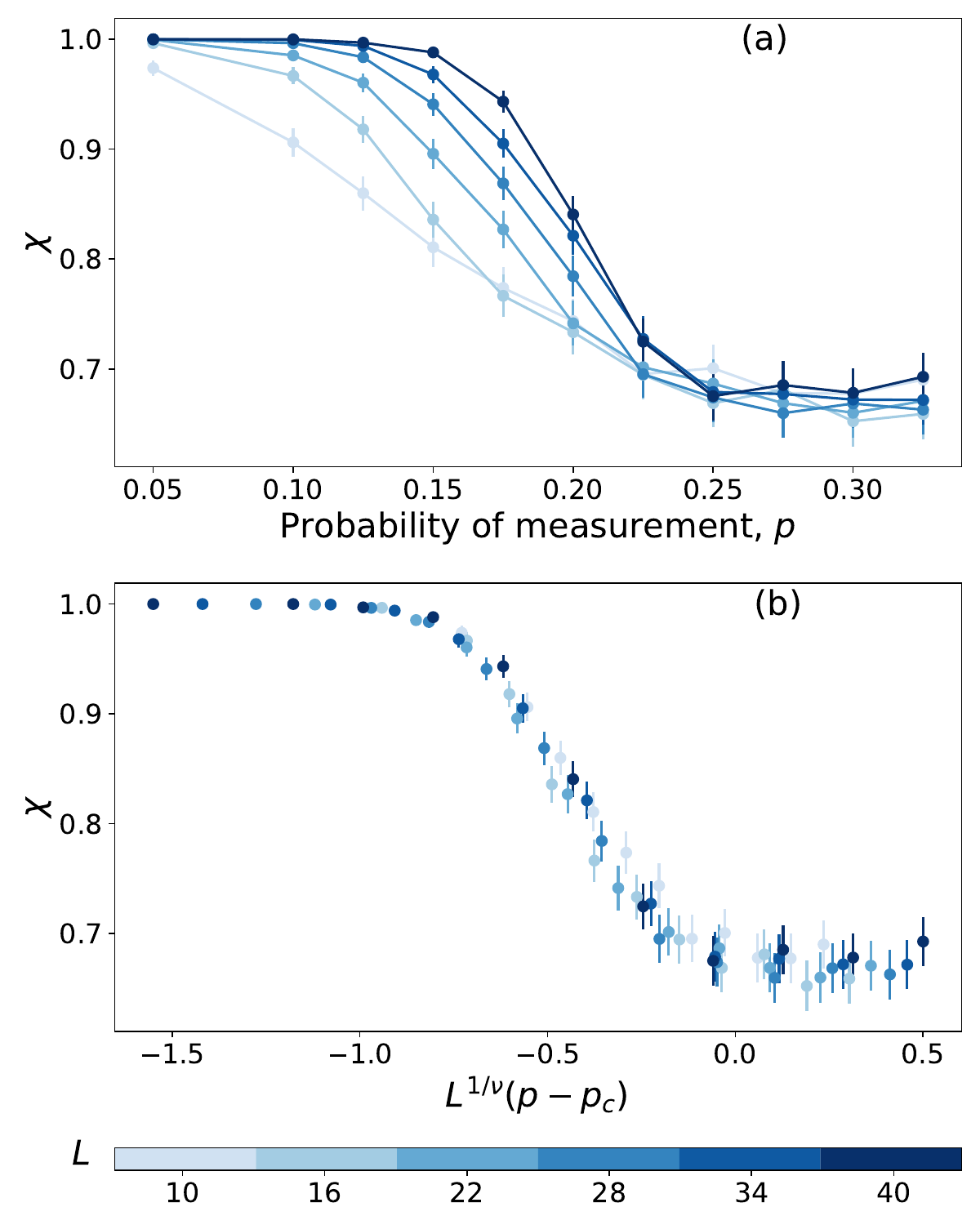}
\caption{(a) Cross entropy $\chi$ for infinite-dimensional systems with up to 20 physical qubits (corresponding to a system size of $L=$40 qubits before compression) computed on \textit{ibm\_sherbrooke}. (b) Collapse of cross entropy curves near the critical point obtained by minimizing the scatter of all points to an unknown scaling function. The fitting procedure gives a critical measurement rate of $p_c=0.26\pm0.02$ and critical exponent $\nu=1.9\pm 0.4$.}
\label{fig:all-to-all-cross-entropy}
\end{figure}

Fig.~\ref{fig:all-to-all-cross-entropy} shows $\chi$ versus $p$ for $p$ increasing from 0.05 to 0.325. The qualitative features of $\chi$ in the all-to-all case are similar to the 1D-chain case, with larger values of $\chi$ for larger systems when $p<p_c$, crossing of all $\chi$ for different $L$ at critical value of $p$, and a plateau to a constant for $p>p_c$. The critical values we extract from fitting to the finite size scaling form Eq.~\eqref{eq:finite-size-scaling} are $p_c=0.26\pm0.02$ and $\nu=1.9\pm 0.4$ at the 90\% confidence level.

A previous mean-field analysis of all-to-all circuits~\cite{nahum2021} predicted $\nu = 2.5$, and numerical simulations of uncompressed noisy circuits in the presence of a 0.1\% erasure channel predicts $\nu \approx 0.8$~\cite{SM}. The presence of noise evidently reduces the value of $\nu$, and may explain why the $\nu$ obtained from experiments is lower than the theoretical value of 2.5. The increased value of $p_c$ for the infinite-dimensional case compared to the 1D case  is consistent with the intuitive picture that entanglement in a system with high connectivity is more stable to measurements that in one in low dimensions.



\textit{Discussion} ---
Our results show that MIPTs can be studied efficiently for systems with various connectivities on near-term superconducting quantum hardware, when restricted to Clifford circuits with an arbitrary initial state.
The cross-entropy protocol used in this Letter eliminates both of the exponential bottlenecks in previous studies of MIPTs on superconducting hardware~\cite{koh2022} while preserving the mid-circuit measurements in the bulk of the circuit, providing a benchmark for the quality of mid-circuit measurements in near-term quantum hardware.
In future work, this protocol may be extended to extract other critical exponents using different circuit architectures~\cite{li2023}; or more broadly, to study various dynamical quantum many-body phenomena~\cite{sang2021measurement, lavasani2021measurement, vasseur2021chargesharpening, tikhanovskaya2023universality, lovas2023quantum, agrawal2023observing}.

\begin{acknowledgements}
H.K. and A.J.M. were supported by the Institute for Quantum Information and Matter and the U.S. Department of Energy under Award No. DE-SC001937. J.S. and A.J.M. were supported by AFOSR under grant number FA9550-23-1-0625.
Y.L. is supported in part by the Gordon and Betty Moore Foundation’s EPiQS Initiative through Grant GBMF8686, and in part by the Stanford Q-FARM Bloch Postdoctoral Fellowship in Quantum Science and Engineering.
M.P.A.F is supported by the Heising-Simons Foundation and the Simons Collaboration on Ultra-Quantum Matter, which is a grant from the Simons Foundation (651457).
Y.L. and M.P.A.F thank Yijian Zou, Paolo Glorioso and Ehud Altman for previous collaborations and discussions.
Y.L. thanks Matteo Ippoliti, Vedika Khemani, and Shengqi Sang for helpful discussions and comments.
M.M. thanks E. Pritchett, S. Sheldon, D. Riste, and P. Rall for useful discussions.

\end{acknowledgements}

\bibliography{references}

\appendix

\section*{End Matter}

\subsection{Resource Analysis}

We now discuss the resource requirements of this protocol compared to previous studies of measurement-induced phase transitions. The demonstrations reported here required fewer than 8 device-hours while retaining the hardware implementation of mid-circuit measurements. In terms of circuit-shots (total number of circuits executed multiplied by the number of shots per circuit), the 1D experiment required a total of $126\times 10^6$ circuit-shots and the all-to-all experiment required $66\times 10^6$ circuit-shots. Both experiments used 1000 shots and 1000 circuits for each $(L, p)$ pair.

The hardware resources required in our experiments are more than two orders of magnitude less than in previous demonstrations \cite{koh2022}. In Ref.~\cite{koh2022}, to obtain the data collapse using tomography required more than $10^{10}$ circuit-shots and only probed system sizes $L \leq 8$.

With the use of circuit compression, we expect that larger systems with as many as  30 physical qubits could be accessed while still maintaining the fidelity of the current experiments.
The limiting factor in the present demonstration up to this scale is only the computational cost of the circuit compression which scales polynomially with system size.
As this work focused on demonstrating the protocol on near-term hardware, we did not emphasize efficient implementations of the classical circuit-compression algorithm; this task could be a focus of future work. To increase to larger system sizes, the bulk and encoding ratios can also be reduced from 3, used in our experiments, to as low as 1 while still maintaining a visible phase transition.


\subsection{Further Discussions on Circuit Compression}

A technique we used throughout this work is Clifford circuit compression, which takes a circuit with non-stabilizer initial states and outputs a gate-efficient representation  (see Supplemental Material for details~\cite{SM}).
Circuit compression allows us to study systems larger than the number of available hardware qubits while minimizing
the number of mid-circuit measurements, which is the slowest element of hybrid circuits.
Circuit compression also makes possible the exploration of related phenomena on graphs that cannot be embedded in 2D, for example on those with all-to-all connectivity~\cite{nahum2020alltoall, sagar2020volume} or on trees~\cite{vasseur2020mft, feng2022measurementinduced}.
From the cross entropy data for $\rho \neq \sigma$ we can extract critical exponents that are evidently comparable to classical simulations and theoretical predictions, even though no error mitigation techniques are applied.
However, circuit compression complicates the propagation of noise, an analysis which we leave for future work.
We note that compression or simplification techniques also exist for non-Clifford circuits~\cite{gujarati2023quantum, debrugiere2024fastershortersynthesishamiltonian}, which we expect to be helpful in extending the XEB protocol beyond the Clifford case.

\subsection{Effects of Noise}

The data for $\chi_{\rho = \sigma}$ in Fig.~\ref{fig:rho=sigma-cross-entropy} shows that the effect of noise increases with the system size, suggesting that stronger error suppression will be needed for further scaling up the system size.
This trend qualitatively agrees with our classical simulations with injected depolarizing noise.
However, the quantitative behavior of $\chi_{\rho = \sigma}$ differs between the experimental data and our classical simulations, suggesting that the noise model based on depolarizing noise is incomplete.
We detail these discussions in the Supplemental Material~\cite{SM}.

A second indication as to the limitations of the depolarizing noise model comes from
comparing $\chi_{\rho = \sigma}$ (Fig.~\ref{fig:rho=sigma-cross-entropy}) with $\chi_{\rho \neq \sigma}$ (Fig.~\ref{fig:1D-chain-cross-entropy}), where we find that the former is often visibly smaller than the latter, particularly for the larger values of $p$ we accessed in our experiments.
On the other hand, as we show in Supplemental Material~\cite{SM} with rigorous analyses, one has the bound $\chi_{\rho = \sigma} \ge \chi_{\rho \neq \sigma}$ in Clifford circuits with a simple noise model, namely those that can be written as stabilizer operations and their probabilistic mixtures.
These include the erasure errors we use in our classical numerical simulations.
We attribute the violation of this bound to real device error, which necessarily involves e.g. coherent and non-unital noise which are not included in our simple noise model.
Evidently, $\chi_{\rho=\sigma}$ is more sensitive to noise than when two different initial states are used.

It will be an interesting future direction to explore the effects and the description of real device noise on the critical properties, and conversely, the extent to which a phase transition in cross entropy can be informative of experimental conditions and changes, such as dynamical decoupling, for large systems where process tomography is too costly.

\subsection{Effects of Error Mitigation}
Improvement of the experimental performance of the processor, for instance by reducing cross-talk and introducing carefully tailored dynamical decoupling sequences may also allow us to explore 
the phase transition in even larger systems. Preliminary experiments including dynamical decoupling show some improvement in the fidelity obtained in the intermediate regime of 5 to 8 hardware qubits; however, for larger systems dynamical decoupling had little effect and so was not used in any of the experiments~\cite{SM}. We also attempted to use readout error mitigation for our $\rho\neq \sigma$ experiments in 1D, but it did not change our results and was not applied to our data~\cite{SM}. Although the two error mitigation methods above do not yield visible improvement on the data quality, we expect that some form of error mitigation (either more sophisticated versions of the above, or other types) will eventually be necessary if the experiments are further scaled up. This is a topic we leave for future work.

\clearpage

\makeatletter
\patchcmd{\frontmatter@abstract@produce}
  {\vskip200\p@\@plus1fil
   \penalty-200\relax
   \vskip-200\p@\@plus-1fil}
  {}
  {}
  {}
\makeatother


\setcounter{equation}{0}
\setcounter{figure}{0}
\setcounter{table}{0}
\renewcommand{\theequation}{S\arabic{equation}}
\renewcommand{\thefigure}{S\arabic{figure}}
\renewcommand{\thesection}{S\arabic{section}}

\widetext

\begin{center}

\textbf{\large Supplemental Materials: Experimental demonstration of scalable cross-entropy benchmarking to detect measurement-induced phase transitions on a superconducting quantum processor}

\end{center}












\tableofcontents

\section{Compression of Clifford circuit with magic initial state} \label{si:compression}
Here we describe the Clifford based compression algorithm we use to reduce the required number of physical qubits by a factor of two, as well as to reduce the total number of mid-circuit measurements to equal the number of physical qubits. The compression is based on Ref. \cite{yoganathan2019} with an improvement that removes the requirement for dynamic circuits (adaptivity), instead using an efficient classical simulation and classical coin flipping.
Here, we first summarize the compression algorithm stated in Ref.~\cite{yoganathan2019}, and then explain how to remove the adaptivity.

\subsection{Summary of the compression algorithm}
In a particular circuit realization the unitaries and the measurements can be written as
\begin{align}
    C_{\mathbf{m}} =  \ldots U_3 M_{m_2} U_2 M_{m_1} U_1.
\end{align}
Here $m_j$ is the $j$-th measurement outcome of the entire record, and correspondingly $M_{m_j} = (1+(-1)^{m_j} P_j)/2$ is the $j$-th projection operator, with $P_j$ the Pauli operator being measured.
Moving all unitaries past the measurements to the right, we can equivalently write
\begin{align}
\label{eq:multi_qubit_meas}
    C_{\mathbf{m}} =  \ldots \widetilde{M}_{m_2} \widetilde{M}_{m_1},
\end{align}
where
\begin{align}
    \widetilde{M}_{m_j} = \frac{1}{2}(1+z_j \widetilde{P_j}), \quad \widetilde{P}_j = U^\dagger_1 U^\dagger_2 \ldots U^\dagger_j P_j U_j U_{j-1} \ldots U_1
\end{align}
are now multi-site Pauli measurements and $z_j=(-1)^{m_j}$.

Let $A = \{1, \dots, k\}$, and $B = \{k+1,\ldots, N\}$.
Following Ref.~\cite{yoganathan2019}, we state without proof that the following algorithm correctly samples an output bitstring of the circuit $C$ on a input state in the new basis, with input states of the form $\ket{\psi} = \ket{\phi_A} \otimes \ket{0^{\otimes N-k}_B}$.
\begin{enumerate}
    \item
    Initialize the quantum state $\ket{\phi_A}$, define the initial stabilizer group $\mathcal{S} =\langle Z_{k+1}, \ldots Z_{N} \rangle$, and let the Pauli operators be $\{ \widetilde{P}_j \}$.
    \item 
    Consider each $\widetilde{P}_{j}$ in increasing order of $j$. For each $j$ there are three possible cases:
    \begin{enumerate}
        \item
        $\widetilde{P}_j \in \mathcal{S}$. In this case the measurement result is deterministic, and can be classically computed and we do not need to update the state or $\mathcal{S}$.
        \item
        $\widetilde{P}_j \notin \mathcal{S}$, and it anticommutes with at least one element $Q \in \mathcal{S}$.
        In this case, the measurement result of $\widetilde{P}_j$ is equally likely $z_j = \pm 1$.
        We can flip a classical coin to sample $z_j$.
        Further, we need to account for the change in the state, which can be shown to be
        \begin{align} 
            \ket{\phi} \to V_j(z_j) \ket{\psi}
        \end{align} 
        where $V_j(z_j)$ is a Clifford unitary operator 
        \begin{align} \label{eq:V}
            V_j(z_j) = \frac{1}{\sqrt{2}} (Q + z_j \widetilde{P}_j).
        \end{align}
        Instead of evolving the state and updating $\mathcal{S}$, we adopt the Heisenberg picture and modify all subsequent measurements $\widetilde{P}_{k>j}$ as follows,
        \begin{align} \label{eq:P_update}
            \widetilde{P}_{k} \to V_j(z_j)^\dagger \widetilde{P}_{k} V_j(z_j), \quad \forall k > j.
        \end{align}
        \item
        $\widetilde{P}_j \notin \mathcal{S}$, and it commutes with all elements of $\mathcal{S}$.
        It then necessarily commutes with $Z_{k+1}, \ldots, Z_{N}$ since these stabilizers are permanent, as we can check at the end of the algorithm (see comment 2 below).
        It follows that $\widetilde{P}_j$ only contains the identity operator or the Pauli $Z$ operator on $B$.
        We can then consider a truncated Pauli operator that is supported only on $A$, 
        \begin{align}
            \widetilde{P}^A_j \coloneqq \eta_j \cdot\widetilde{P}_j|_A,
        \end{align}
        where 
        $\widetilde{P}_j|_A$ is the restriction of $\widetilde{P}_j$ on $A$, and the sign
        $\eta_j = \pm 1$ can be chosen such that for any state $\ket{\phi_A}$ we have
        \begin{align}
            \bra{\phi_A} \widetilde{P}^A_j \ket{\phi_A} = \bra{\phi_A \otimes 0^{\otimes N-k}_B} \widetilde{P}_j \ket{\phi_A \otimes 0^{\otimes N-k}_B}.
        \end{align}
        The measurement of $\widetilde{P}_j$ on the joint system $AB$ can therefore be faithfully simulated by a measurement of $\widetilde{P}^A_j$ on just $A$.
        We perform this measurement on the state $\ket{\phi_A}$, update the state accordingly and record the measurement result $z_j^\prime$.
        We then update the stabilizer group as
        \begin{align} 
            \mathcal{S} \to \langle \mathcal{S}, z_j^\prime \widetilde{P}^A_j \rangle.
        \end{align}
    \end{enumerate}
\end{enumerate}

We see that in this algorithm
\begin{enumerate}
    \item Cases (1) and (2) can be accounted for by classical simulation, and only in case (3) a quantum operation on $\ket{\phi_A}$ needs to be performed.
    \item The stabilizer group $\mathcal{S}$ gets augmented only in case (3), and can be augmented at most $k$ times.
    Once an operator is added into $\mathcal{S}$, it will remain in $\mathcal{S}$ until the algorithm terminates.
\end{enumerate}
In this way, a given sequence of multi-site measurements can be simulated by a ``compressed circuit'' with at most $k$ multi-site measurements on $A$, as well as classical coin flips, up to a polynomial time overhead.


\subsection{Removal of adaptivity}
A technical problem of the above algorithm is that the update of the stabilizer group $\mathcal{S}$ in case (c) depends on the quantum measurement result $z_j^\prime$.
Not knowing $z_j^\prime$ before the circuit execution will lead to the lack of knowledge of the sign of $Q\in\mathcal{S}$ in case (b) if occuring after the update of $\mathcal{S}$ due to case (c).
Here we show the adaptivity can be removed by proving that the effect of flipping signs of $z_j^\prime$ or $Q$ can be captured by classical postprocessing.

In order to prove it, we first notice that $Q\rightarrow -Q$ is equivalent to $z_j\rightarrow -z_j$ in Eq.~\ref{eq:V} ($V \rightarrow -V$ has no effect on Eq.~\ref{eq:P_update}).
We additionally notice that 
\begin{align}
    V_j(-z_j) = Q V_j(z_j) Q,
\end{align}
so that for any $k > j$,
\begin{align}
\label{eq:adativity}
    V_j(-z_j)^\dagger \widetilde{P}_k V_j(-z_j) &= Q V_j(z_j)^\dagger Q \widetilde{P}_k Q V_j(z_j) Q \nonumber\\ 
    &= \lambda_{Q, \widetilde{P}_k} Q V_j(z_j)^\dagger \widetilde{P}_k V_j(z_j) Q \nonumber\\
    &= \lambda_{Q, \widetilde{P}_k} \lambda_{Q,  V_j(z_j)^\dagger \widetilde{P}_k V_j(z_j)} V_j(z_j)^\dagger \widetilde{P}_k V_j(z_j),
\end{align}
where we have defined the commutator of Pauli operators $A, B$
\begin{align}
    AB = \lambda_{A,B} BA.
\end{align}
Eq.~\eqref{eq:adativity} implies that flipping measurement results $z_j^\prime$ at most result in sign changes of the subsequent measurements operators $\widetilde{P}_{k>j}$, and such sign dependence can be classically captured.
In practice, we can first determine the form of each Pauli operator to be measured on $A$ in the compressed circuit, and assume they all have +1 sign; the adaptivity can be re-introduced in post-processing, by flipping the measurement results appropriately.

\subsection{Decomposition of the Pauli-based computing model to a common gate set} \label{si:PBC_to_gate}


Here we describe an algorithm to decompose each multi-qubit Pauli measurement in Eq.~\eqref{eq:multi_qubit_meas}
to
\[P_j=\left(\prod_i C_i\right)^\dagger Z[k] \left(\prod_i C_i\right),\]
where $\{C_i\}$ contains up to $m$ single-qubit Clifford operations and $2m$ CNOT gates.
For a Pauli string $P_j=\otimes_{i=1}^{m} P_j[i]$, where $P_j[i]\in{I,X,Y,Z}$, we first convert each $X$ and $Y$ to a Pauli $Z$ at qubit $i$ by a single-qubit Clifford operation $C[i]$, i.e. $C[i] P_j[i] C[i]^\dagger = Z[i]$. After this step, the Pauli string becomes a string of $I$s and $Z$s.
We note the fact that $\text{CNOT}_{1,2}(I\otimes Z) \text{CNOT}_{1,2} = (Z\otimes Z)$ and $\text{CNOT}_{1,2}(Z\otimes Z) \text{CNOT}_{1,2} = (I\otimes Z)$.
Thus we first sequentially convert the Pauli string to the form of $I...IZ...ZI...I$ by converting adjacent $ZI$ or $IZ$ to $ZZ$, and then sequentially convert it to $I...IZI...I$ with a single $Z$ in the middle by converting adjacent $ZZ$ to $IZ$ or $ZI$.

By using the above algorithm for the decomposition of $P_{j=1}$, we obtain $P_{j=1}=(\prod C_i)^\dagger Z[k] (\prod C_i)$.
However, instead of naively applying the algorithm for each $P_{j}$, we first ``absorb'' $(\prod C_i)^\dagger$ into the rest of the Pauli strings by $P_{j}\rightarrow (\prod C_i)P_{j}(\prod C_i)^\dagger$, and then apply the above algorithm to the next Pauli measurement. By doing such ``absorption", we roughly reduce the number of CNOT gates by half.
Finally, 
the compressed circuit is decomposed to at most $m^2$ single-qubit gates and $2m^2$ CNOT gates.


\subsection{Resource reduction after circuit compression}\label{si:compressed_circuit_resources}


\begin{table}[!hbt]
    \centering
    \begin{tabular}{|c|c|c|c|}
    \hline
    & 1D circuits, uncompressed & All-to-all circuits, uncompressed & 1D and all-to-all circuits, compressed\\
    \hline
    Num. hardware qubits & $L$ & $L$ & $L/2$\\
    \hline
    Average depth & $6L$ & $2L^3$ & $L^2$\\
    \hline
    Avg. num. 2 qubit gates & $3L^2$ & $2L^3 + 3L^2$ & $L^2/2$\\
    \hline
    Avg. num. measurements & $3L^2p$ & $3L^2p$ & $L/2$\\
    \hline
    \end{tabular}
    \caption{Hardware resources required before and after Clifford circuit compression for a fixed $L$ and $p$. The number of hardware qubits, average depth, and average number of 2 qubit gates required are reduced by a constant factor after compression, whereas the average number of measurements is reduced by a factor of $L$ and is independent of $p$. The values in this table apply both to the 1D system as well as the all-to-all system.}
    \label{table:compression_resource}
\end{table}

In Table \ref{table:compression_resource} we present a summary of the quantum hardware resource requirements before and after circuit compression. Here we are setting $t_\mathrm{bulk}=t_\mathrm{encoding}=3L$ and using an initial $\rho$ state that is an alternating magic state. The number of hardware qubits as well as the number of 2 qubit gates are both reduced by a constant factor after compression, while the average number of measurements becomes independent of the measurement rate $p$. For both circuits, we assume that the hardware topology is linearly arranged qubits with nearest neighbor interactions. For the all-to-all circuits, we therefore require SWAP gates to implement the 2 qubit interactions, resulting in a larger average depth as well as more 2 qubit gates than for 1D circuits. 

We note that although the depth increases by a factor of $L$ after circuit compression for the 1D circuits, for the system sizes in our experiments the increase in depth was not a limiting factor.

\section{Noise model and classical noisy simulations} \label{si:simulated_data}

In this section, we provide classical numerical simulations as a reference for experimental data presented in the main text.
All circuits considered here are drawn from the same ensemble as the experimental runs, and are simulated without compression.

\subsection{1D circuits \label{sec:1D-noisy}}

\begin{figure}
    \centering
    \includegraphics[width=0.32\linewidth]{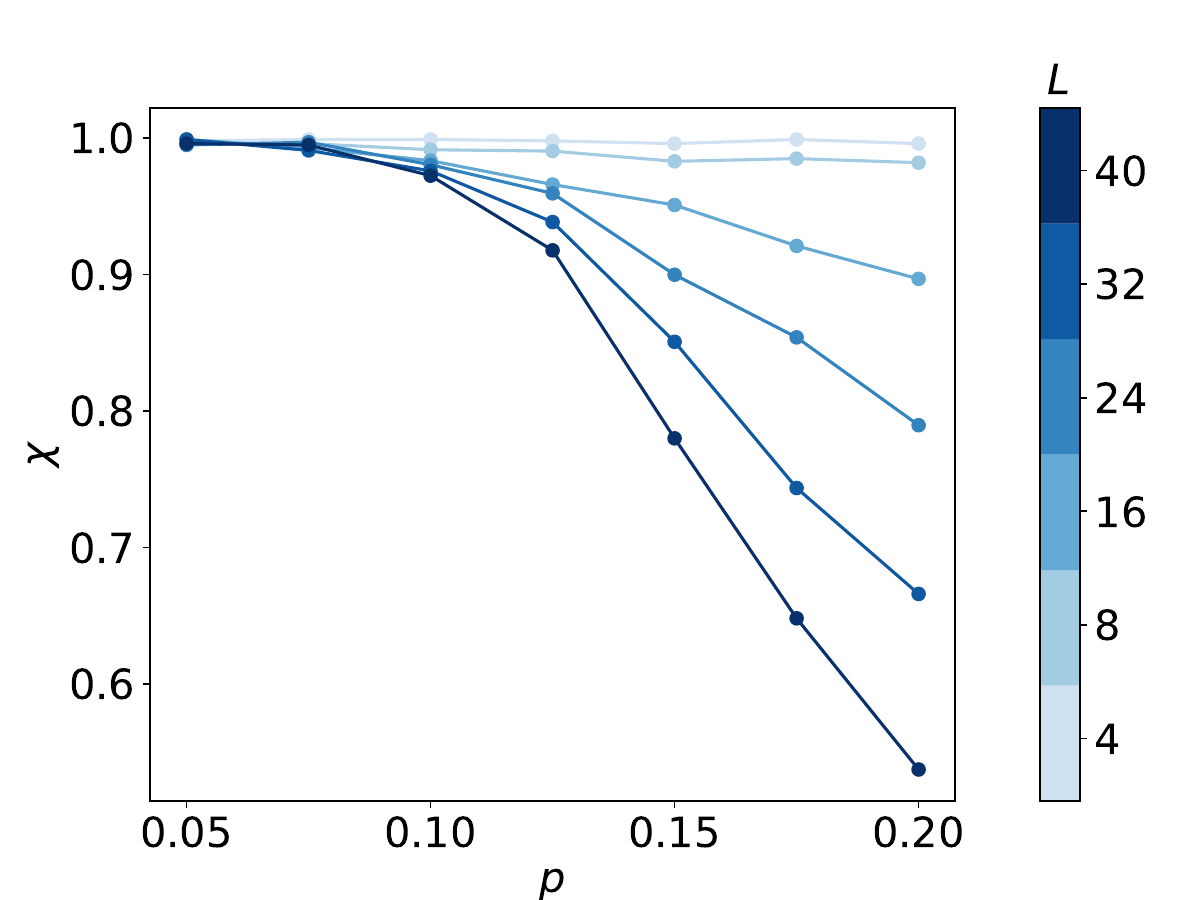}
    \includegraphics[width=0.32\linewidth]{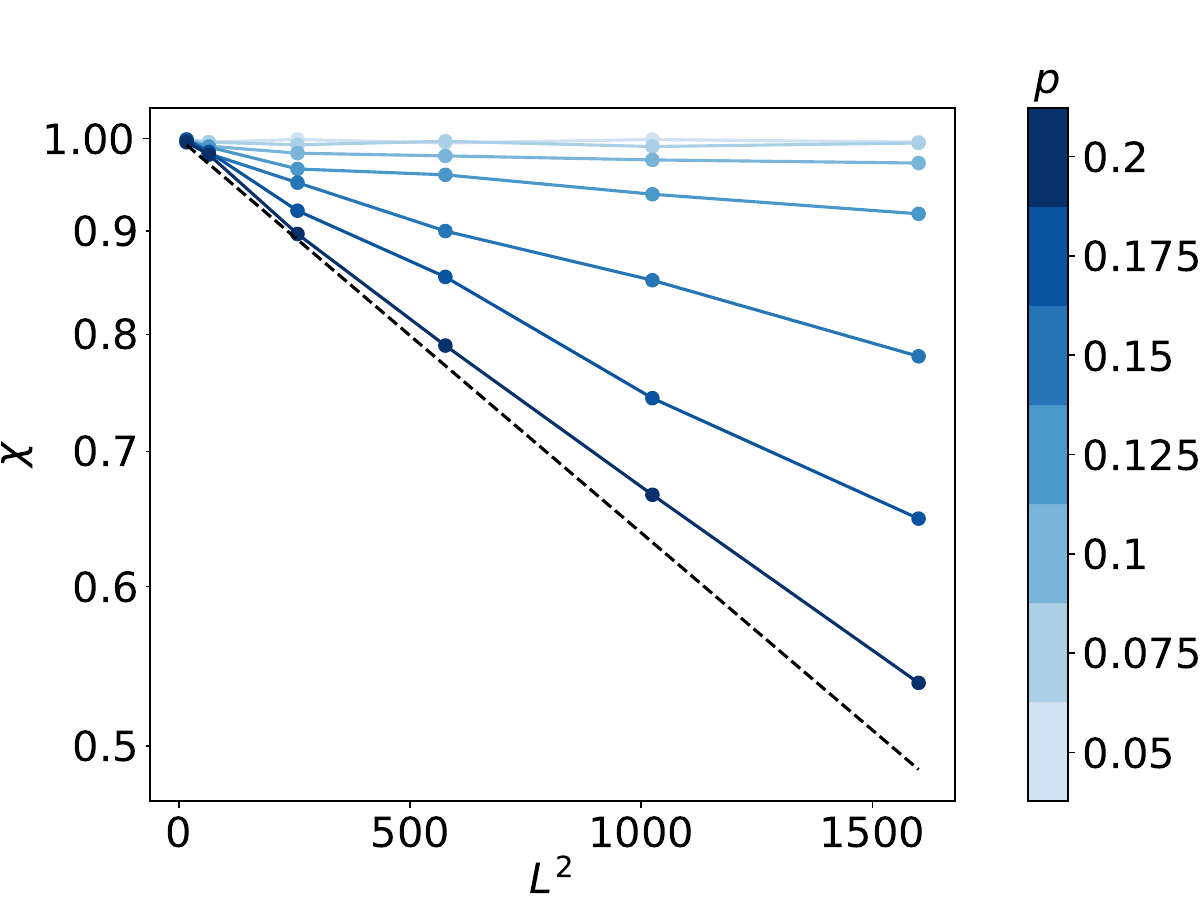}
    \includegraphics[width=0.32\linewidth]{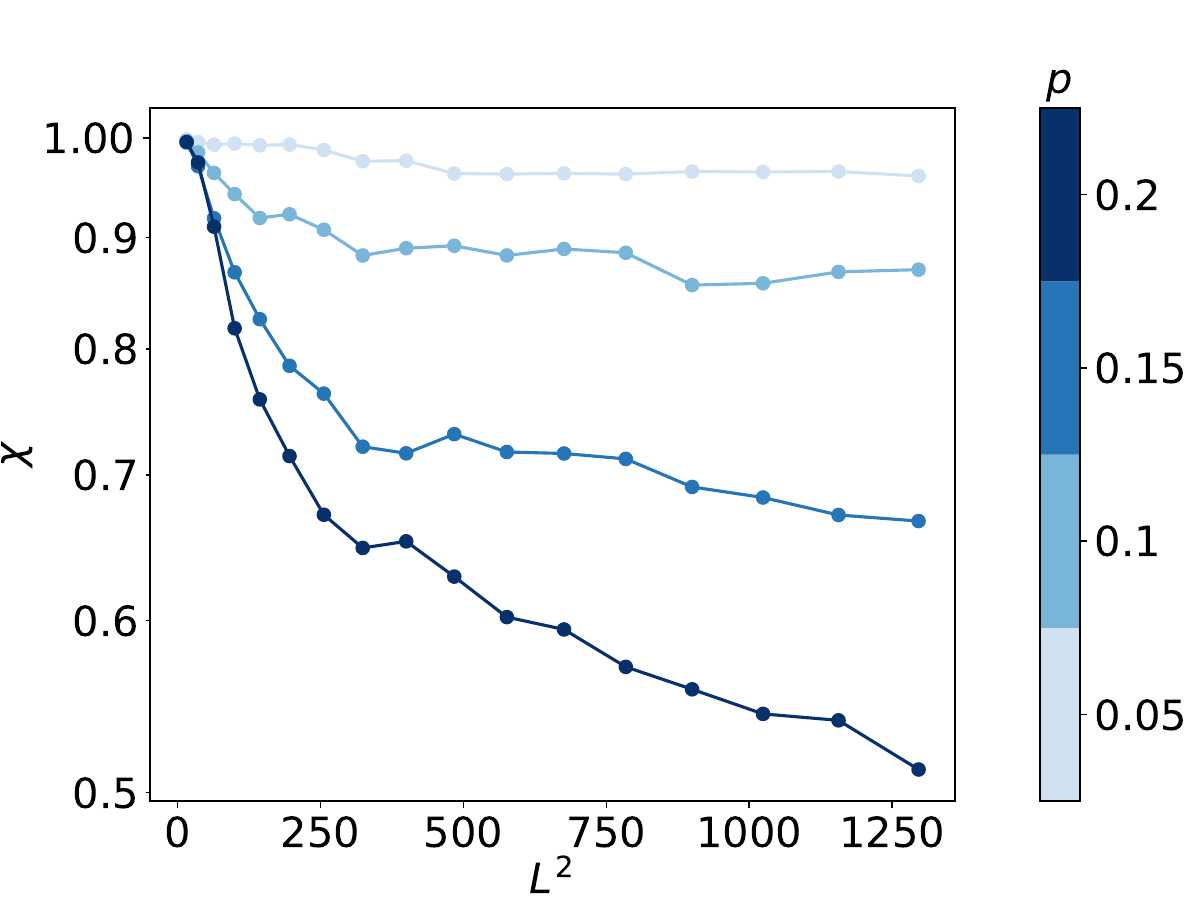}
    \caption{
    (Left) Results from noisy numerical simulations of Clifford circuits in 1D, for system sizes $L \leq 40$.
    We take the initial states $\rho = \sigma = (\ket{0}\bra{0})^{\otimes L}$ as in Fig.~\ref{fig:rho=sigma-cross-entropy}, and randomly insert an erasure channel at each spacetime location of the $\rho$-circuit with probability $q = 0.1\%$.
    (Middle) We find the data consistent with the functional form in Eq.\eqref{eq:noisy_rho=sigma_functional_form}, as indicated by the dashed line.
    (Right) Experimentally obtained $\chi$.
    The non-linear behaviour may be caused due to coherent errors or other noise sources not captured by an erasure channel. 
    }
    \label{fig:1D-simulation-noisy-rho=sigma}
\end{figure}

For the 1D case, we first choose $\rho = \sigma = (\ket{0}\bra{0})^{\otimes L}$, as in Fig.~\ref{fig:rho=sigma-cross-entropy}.
We model the noisy circuit by inserting an erasure channel at each spacetime location of the $\rho$-circuit with probability $q = 0.1\%$, while keeping the $\sigma$-circuit noiseless. This value of $q$ is compatible with values used in other works modeling near-term devices as described in the main text.
The erasure channel replaces the local density matrix with a maximally mixed one, and upon averaging over random circuit realizations becomes a weak depolarizing channel,
\begin{equation}
    \mathcal{E}_x(\rho) = (1-q) \rho + q \left[ \left(\frac{\mathbb{1}}{2}\right)_x \otimes \tr_x \rho \right].
\end{equation}
The results are shown in Fig.~\ref{fig:1D-simulation-noisy-rho=sigma}(a), where we see a decrease in $\chi^{\rm noisy}_{\rho = \sigma}$ when either $L$ or $p$ is increased.
This trend is \textit{qualitatively} consistent with that in Fig.~\ref{fig:rho=sigma-cross-entropy}.

\begin{figure}[h!]
    \centering
    \includegraphics[width=0.45\linewidth]{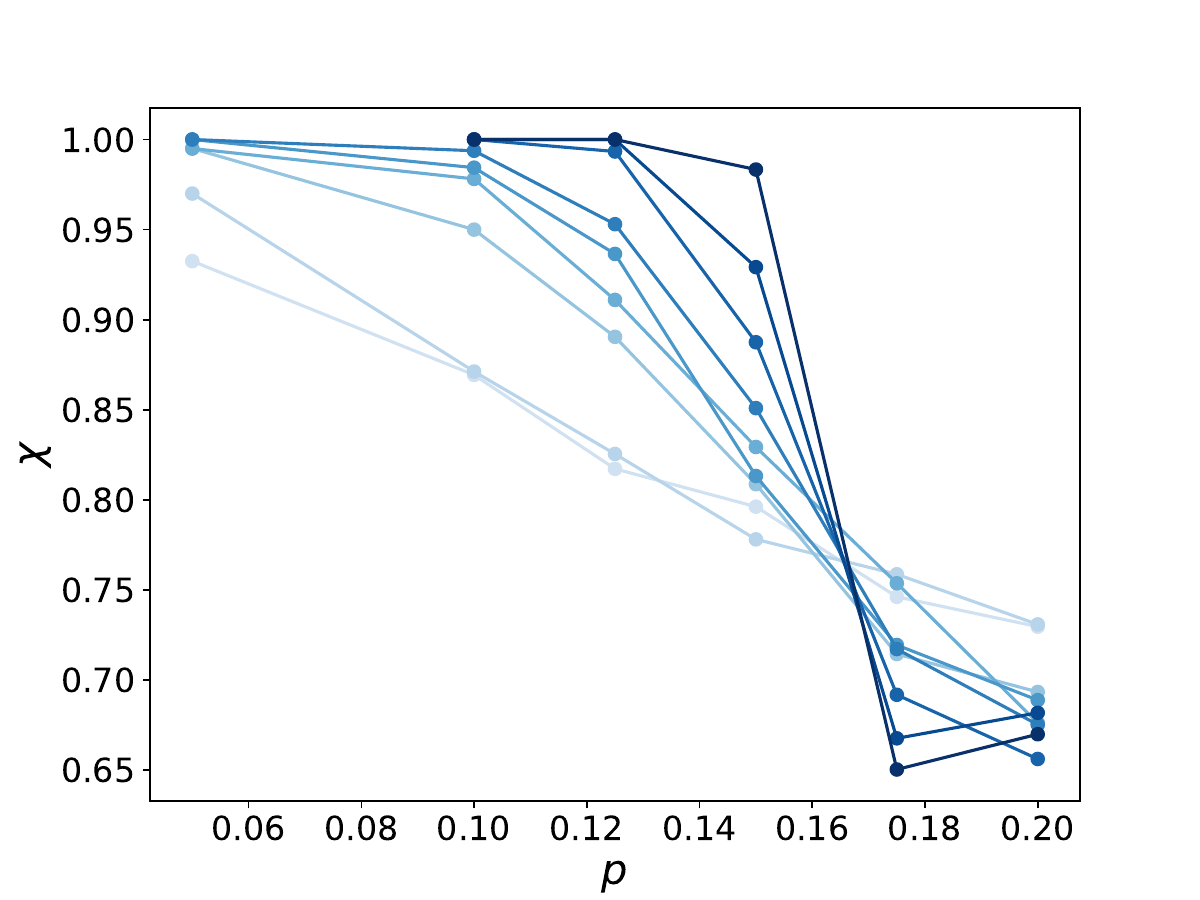}
    \includegraphics[width=0.45\linewidth]{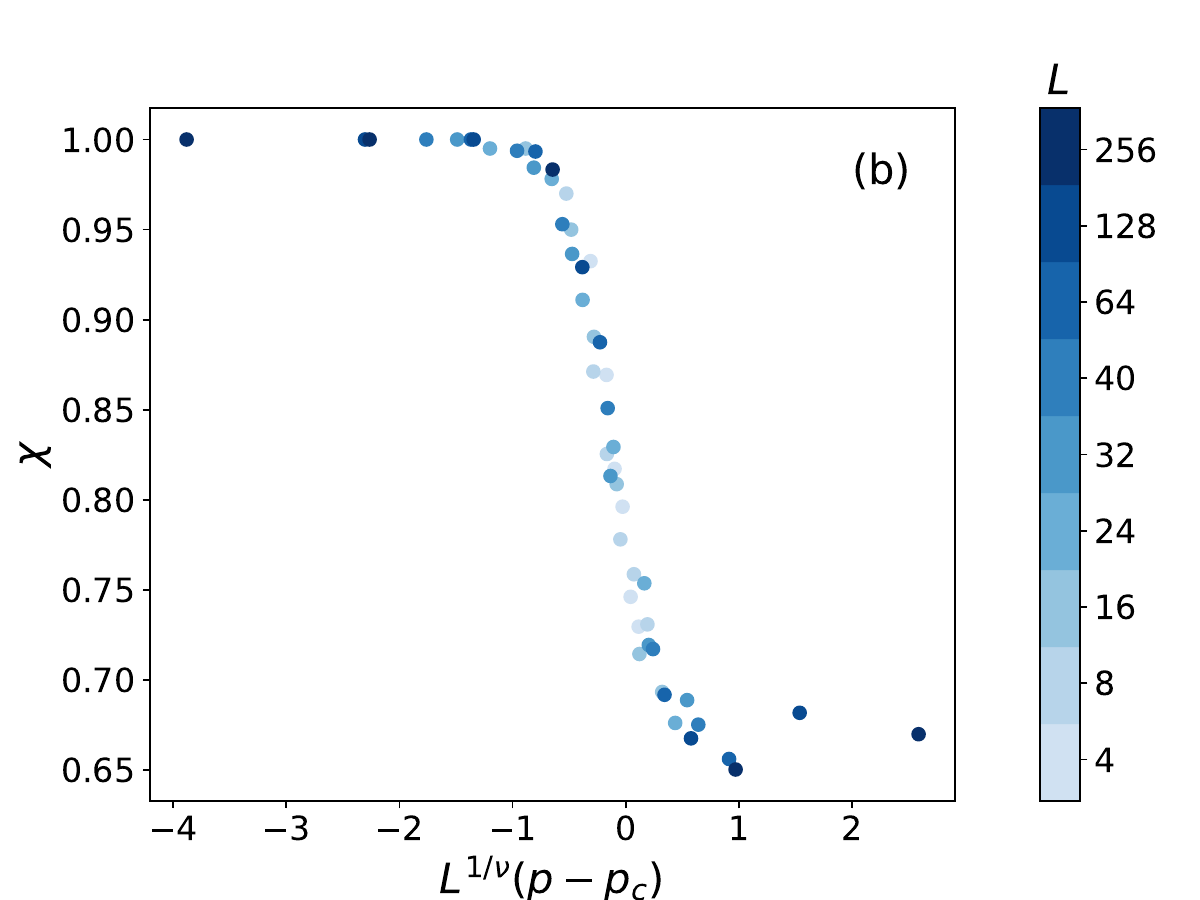}
    \caption{
    (Left) Results from noiseless numerical simulations of Clifford circuits in 1D, for system sizes $L \leq 256$.
    In our simulation, we take  $\rho = \frac{1}{2^L} \mathbb{1}$ and $\sigma = (\ket{0}\bra{0})^{\otimes L}$, as in Ref.~\cite{li2023}.
    (Right) When fitting the data to the scaling form in Eq.~\eqref{eq:finite-size-scaling}, we obtain $p_c = 0.16 \pm 0.01$ and $\nu = 1.3 \pm 0.1$, as consistent with Ref.~\cite{li2023}.}
    \label{fig:1D-simulated-noiseless}
\end{figure}

Next, we consider the $\rho \neq \sigma$ case, but instead with stabilizer initial states $\rho = \frac{1}{2^L} \mathbb{1}$ and $\sigma = (\ket{0}\bra{0})^{\otimes L}$ to facilitate efficient classical simulation.
In Fig.~\ref{fig:1D-simulated-noiseless}(a), we present numerical results obtained from a noiseless simulation.
The overall trend of the results are in qualitative agreement with those in Fig.~\ref{fig:1D-chain-cross-entropy}.
The data collapse in Fig.~\ref{fig:1D-simulated-noiseless}(b) is performed with $p_c = 0.16$ and $\nu = 1.33$, as consistent with Ref.~\cite{li2023}.

\begin{figure}
    \centering
    \includegraphics[width=0.45\linewidth]{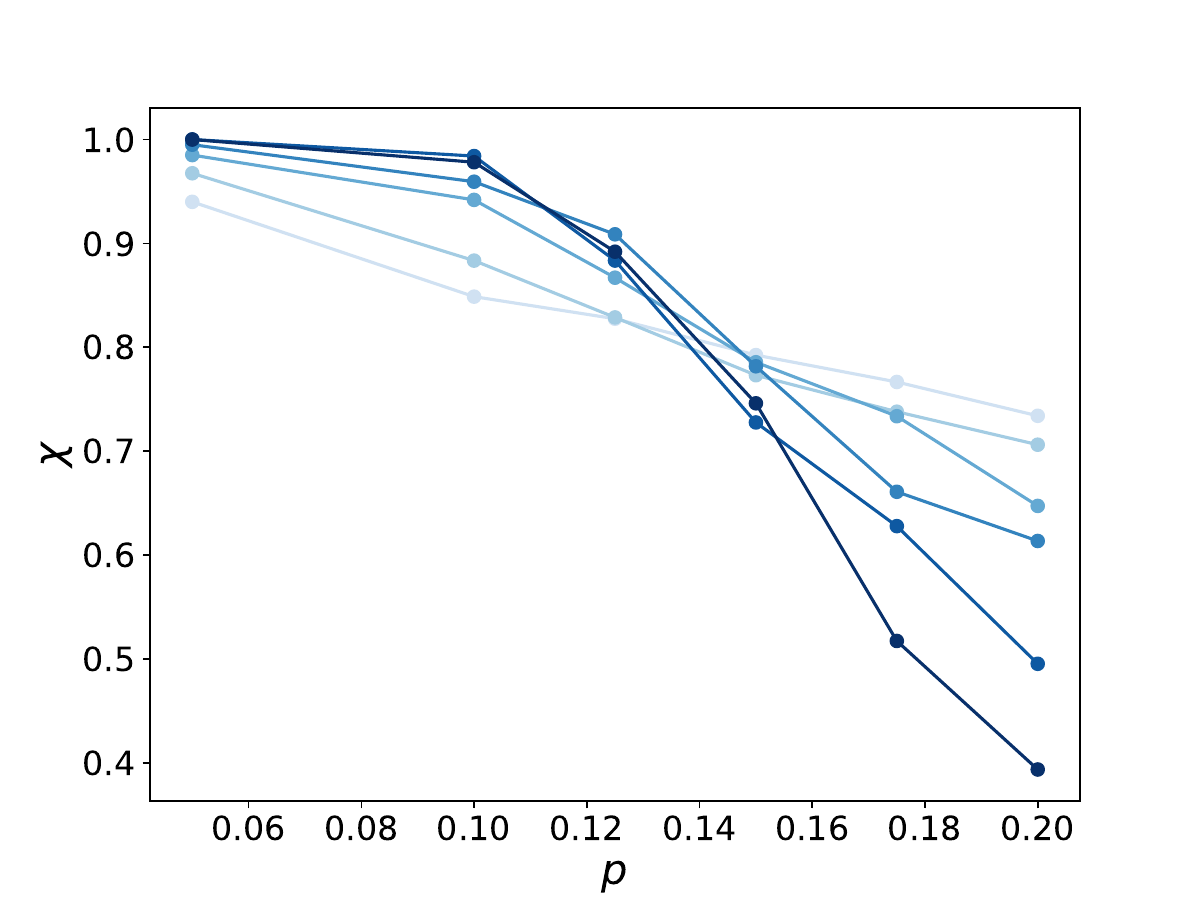}
    \includegraphics[width=0.45\linewidth]{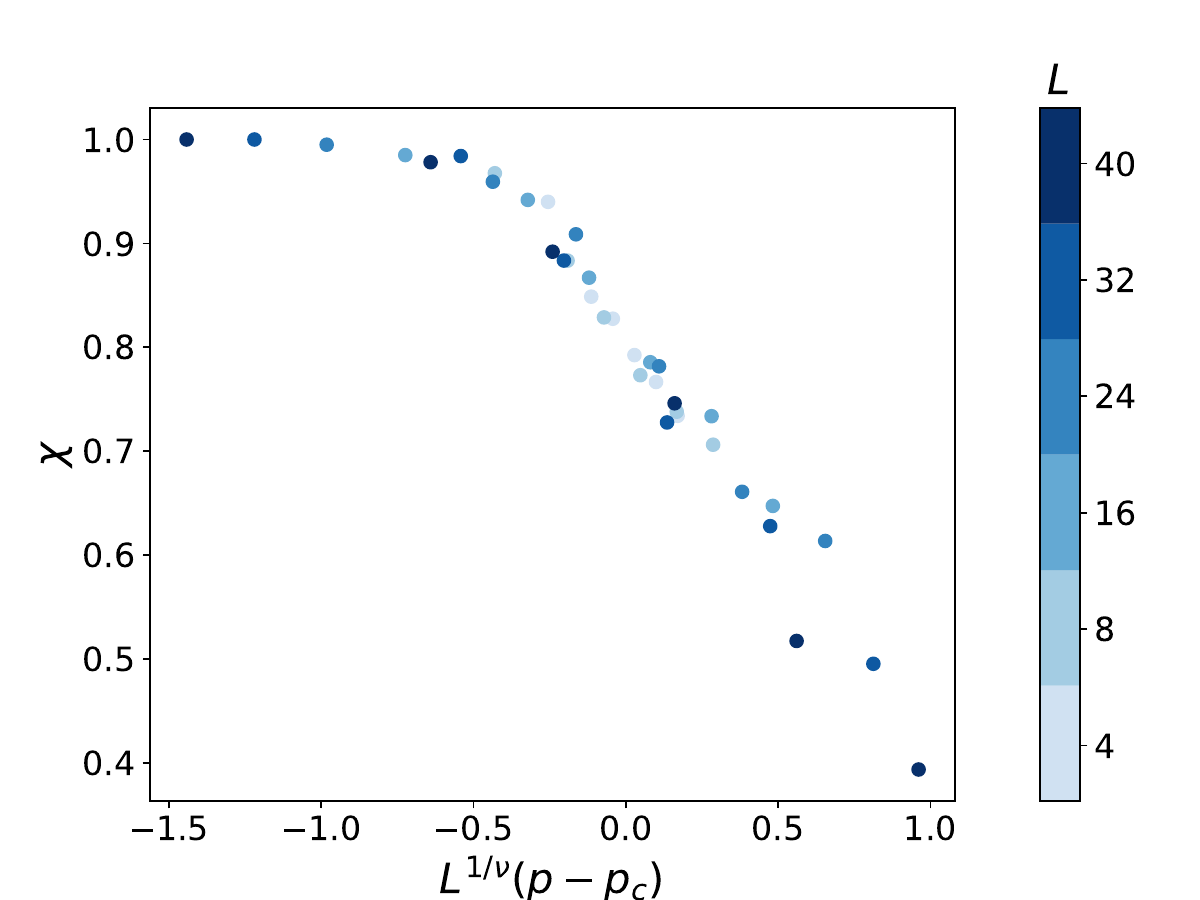}
        \caption{
    (Left) Results from noisy numerical simulations of Clifford circuits in 1D, for system sizes $L \leq 40$.
    We take the same initial states $\rho$ and $\sigma$ as in Fig.~\ref{fig:1D-simulated-noiseless}, and randomly insert an erasure channel at each spacetime location of the $\rho$-circuit with probability $q = 0.1\%$.
    (Right) When fitting the data to the scaling form in Eq.~\eqref{eq:finite-size-scaling}, we use $p_c = 0.14 \pm 0.01$ and $\nu = 1.33 \pm 0.5$ as obtained from Fig.~\ref{fig:1D-chain-cross-entropy} in the main text, where we find consistency.}
    \label{fig:1D-simulated-noisy}
\end{figure}

We also perform a noisy simulation for  $\rho = \frac{1}{2^L} \mathbb{1}$ and $\sigma = (\ket{0}\bra{0})^{\otimes L}$, where we insert an erasure channel at each spacetime location of the $\rho$-circuit with probability $q = 0.1\%$.
The numerical results are shown in Fig.~\ref{fig:1D-simulated-noisy}.
As we anticipate from statistical mechanics arguments (see Ref.~\cite{li2023} and below), for any finite noise rate, the cross entropy will be suppressed to zero for all value of $p$, in the thermodynamic limit.
For small system sizes (before the cross entropy is reduced to zero) the curves will instead appear to cross at a smaller value of $p_c$.
Indeed, the best fit for $p_c$ has now shifted to a smaller value, $p_c \approx 0.14$ (whereas we use the same value for $\nu$), close to the one used for fitting in the main text.

The classical simulation data can be fitted to the following functional form,
\begin{align}
\label{eq:noisy_rho=sigma_functional_form}
    \chi^{\rm noisy}_{\rho = \sigma} \propto \exp \left[-\alpha(p,q) \cdot L^2\right], 
\end{align}
where $\alpha(p,q)$ is a nonzero coefficient depending on $p$ and $q$, see Fig.~\ref{fig:1D-simulation-noisy-rho=sigma}(b).

As we will explain below, this form can be motivated from a statistical mechanics picture, see Eq.~\eqref{eq:chi_rho=sigma_noisy_Ising}.
However, this functional form is not in full agreement with the experimental data, see Fig.~\ref{fig:1D-simulation-noisy-rho=sigma}(c), suggesting that some features of the experimental noise cannot be captured by this simple model. Nonetheless, the qualitative trends are compatible.

\subsection{All-to-all circuits}

\begin{figure}
    \centering
    \includegraphics[width=0.8\linewidth]{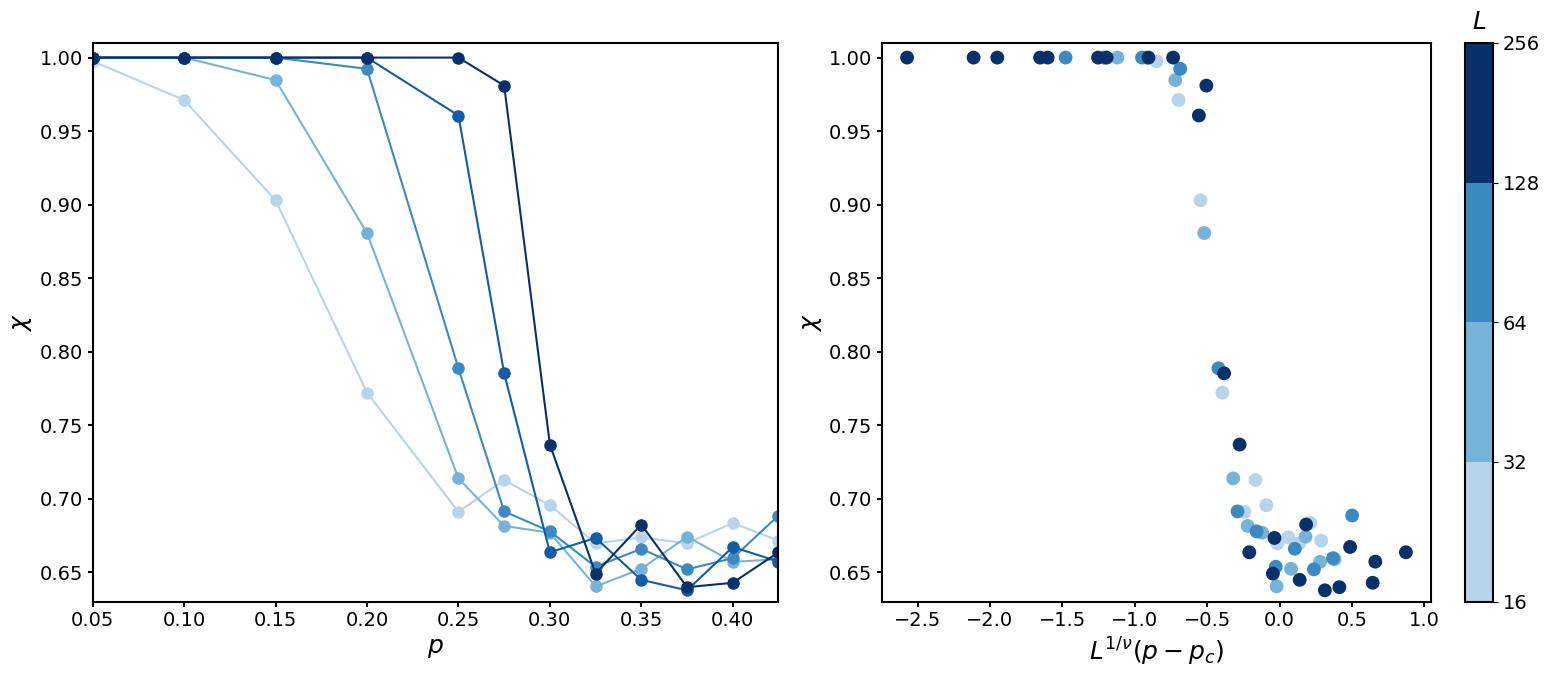}
    \caption{
    (Left) Results from noiseless numerical simulations of Clifford circuits with all-to-all connectivity, for system sizes $L \leq 256$.
    In our simulation, we take  $\rho = \frac{1}{2^L} \mathbb{1}$ and $\sigma = (\ket{0}\bra{0})^{\otimes L}$, identical to our choices in Fig.~\ref{fig:1D-simulated-noiseless}.
    (Right) When fitting the data to the scaling form in Eq.~\eqref{eq:finite-size-scaling}, we obtain $p_c = 0.33 \pm 0.02$ and $\nu = 2.50 \pm 0.2$.}
    \label{fig:all-to-all-q-0.0-L-large}
\end{figure}

\begin{figure}
    \centering
    \includegraphics[width=\linewidth]{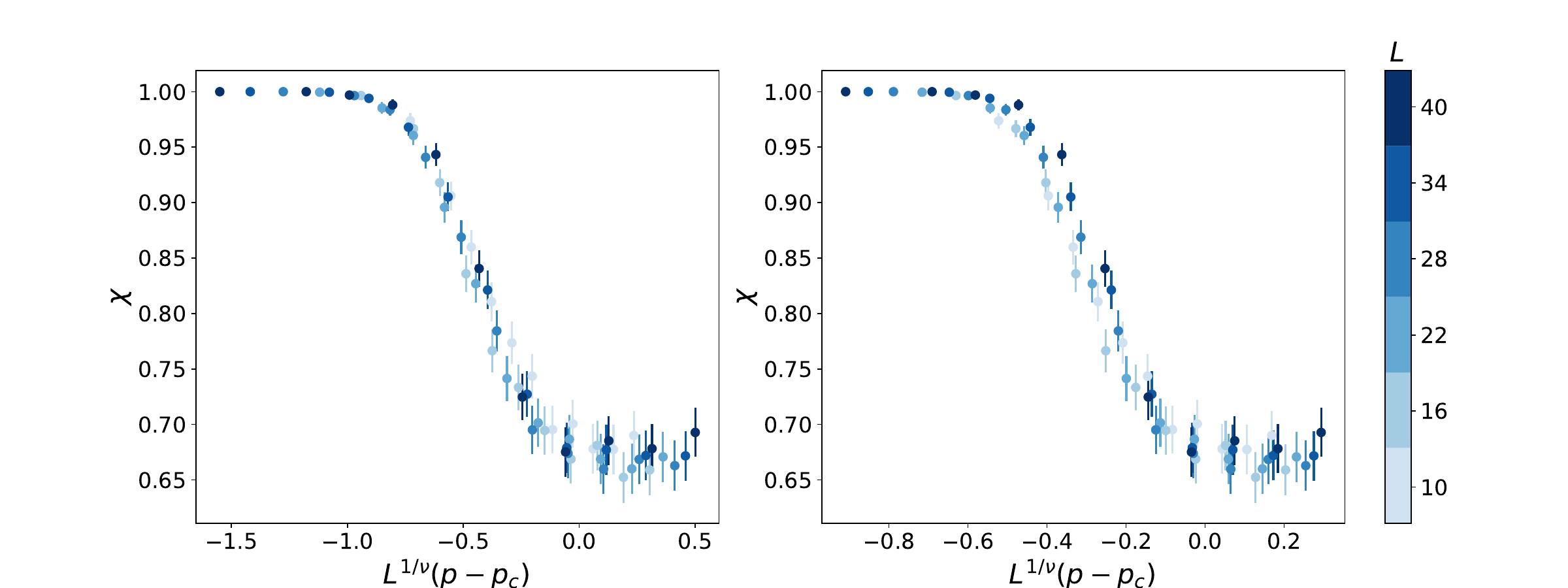}
    \caption{(Left) Collapsing the experimental cross entropy data when fitting both $\nu$ and $p_c$ following the procedures described in Sec.~\ref{si:collapse}.
    This procedure yields $p_c = 0.26 \pm 0.02$ and $\nu = 1.9 \pm 0.4$.
    This plot is reproduced in the main text as Fig.~\ref{fig:all-to-all-cross-entropy}.
    (Right) Data collapse for the same experimental data when setting $\nu$ to its theoretical value of $\nu=2.5$, while keeping $p_c = 0.26$ (the same as in the left panel).}
    \label{fig:all-to-all-nu_theory}
\end{figure}

\begin{figure}
    \centering
    \includegraphics[height=0.24\linewidth]{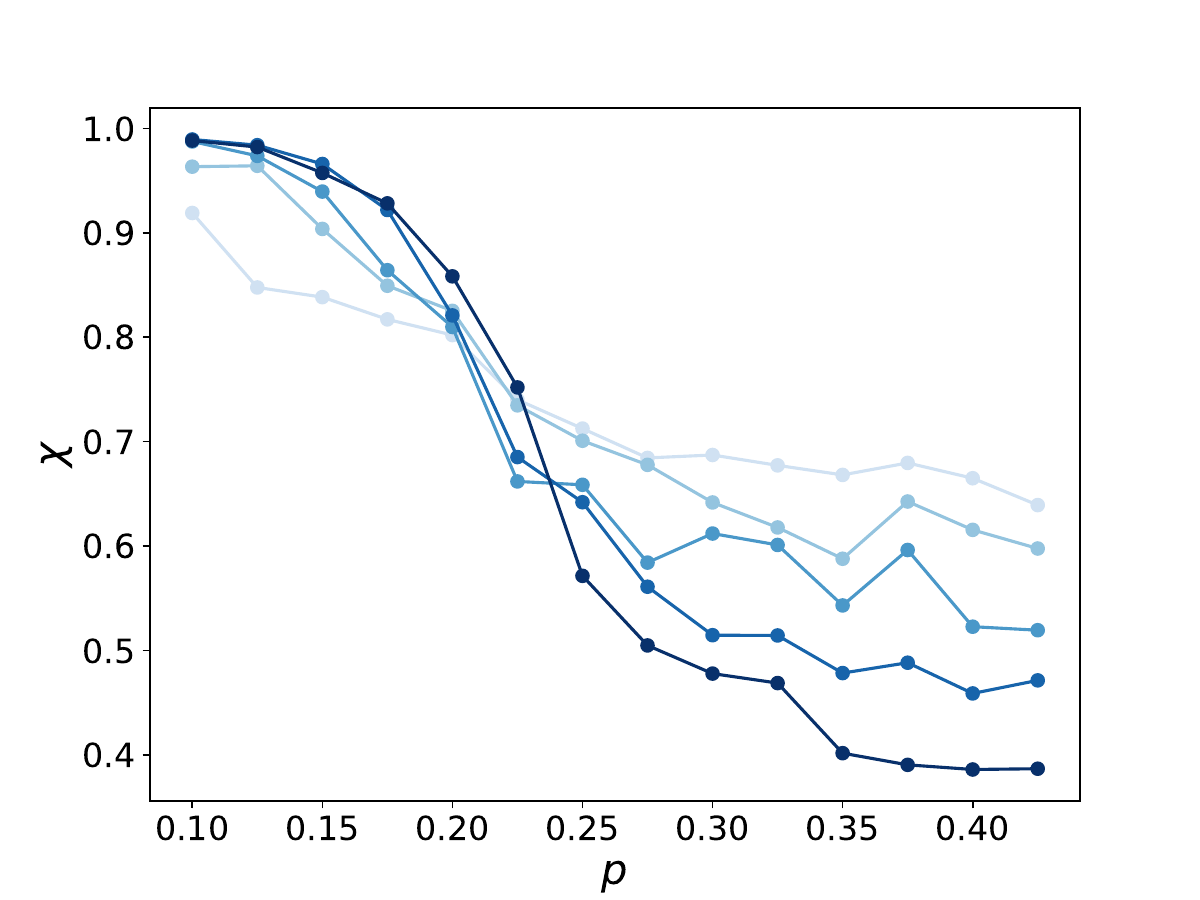}
    \includegraphics[height=0.24\linewidth]{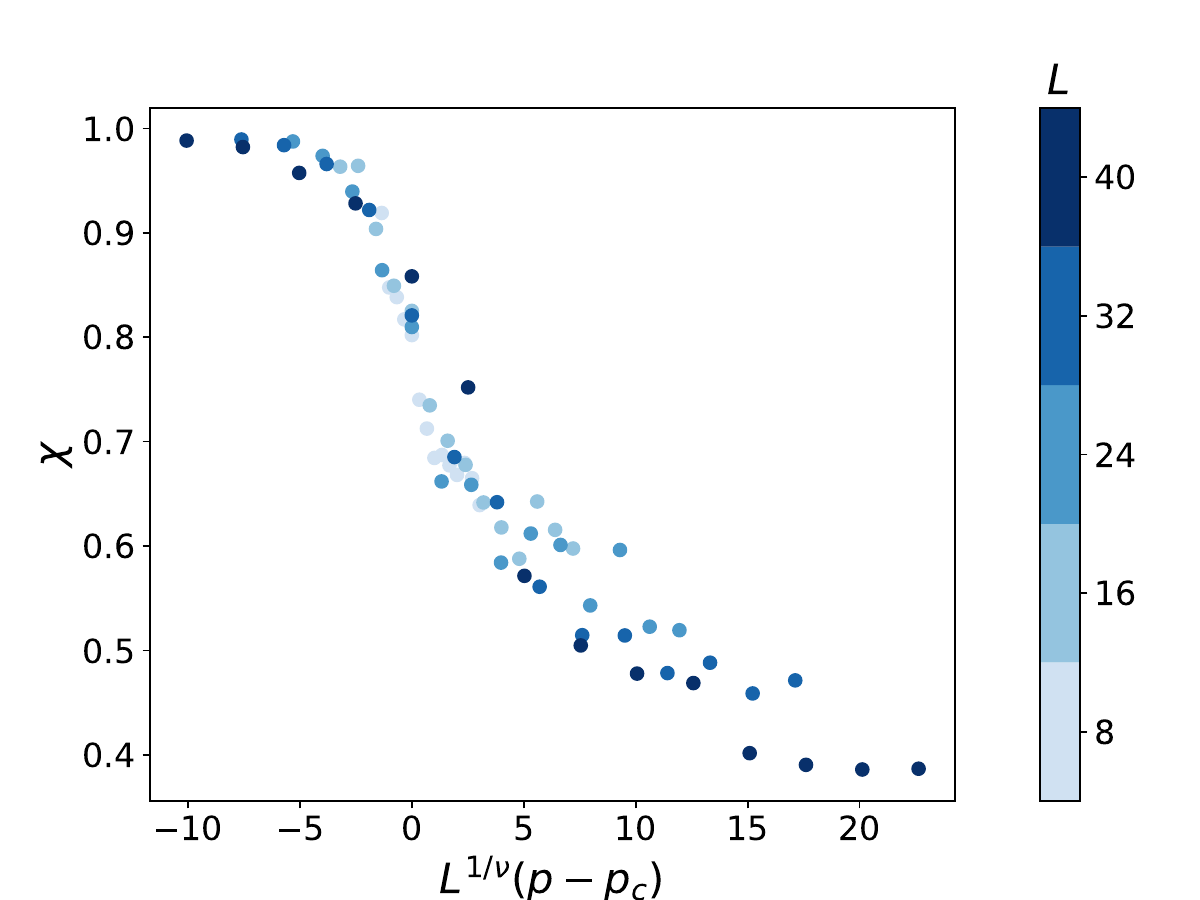}
    \includegraphics[height=0.22\linewidth]{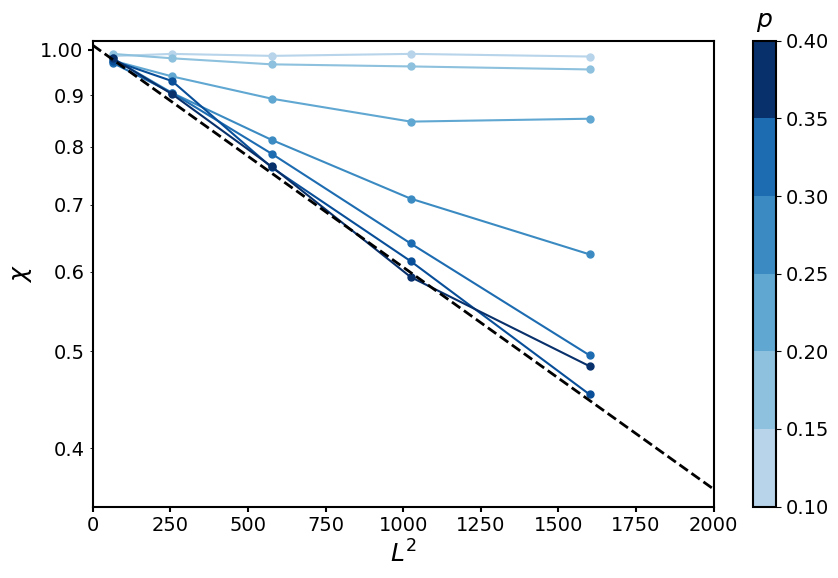}
    \caption{
    (Left) Results from noisy numerical simulations of Clifford circuits with all-to-all connectivity, for system sizes $L \leq 40$.
    We take the same initial states $\rho$ and $\sigma$ as in Fig.~\ref{fig:all-to-all-q-0.0-L-large}, and randomly insert an erasure channel at each spacetime location of the $\rho$-circuit with probability $0.1\%$.
    (Middle) When fitting the data to the scaling form in Eq.~\eqref{eq:finite-size-scaling}, we find $p_c = 0.20 \pm 0.02$ and $\nu = 0.80 \pm 0.3$.
    (Right) For $\rho = \sigma = \frac{1}{2^L} \mathbb{1}$ in the all-to-all circuit simulation with random erasure errors occurring with probability $0.1\%$, the data for $\chi_{\rho = \sigma}^{\rm noisy}$ can again be fit to the functional form in Eq.~\eqref{eq:noisy_rho=sigma_functional_form}, as highlighted by the dashed line.
    }
    \label{fig:all-to-all-q-0.001-L-small}
\end{figure}


We also performed classical numerical simulations for circuits with all-to-all connectivity, taking the same initial states as our 1D simulations.
The results are shown in Fig.~\ref{fig:all-to-all-q-0.0-L-large},\ref{fig:all-to-all-q-0.001-L-small}. From the noiseless simulation (Fig.~\ref{fig:all-to-all-q-0.0-L-large}) of $L \leq 256$ we obtain fits $p_c = 0.33 \pm 0.02$ and $\nu =  2.50 \pm 0.2$. 
In particular, the critical exponent $\nu \approx 2.50$ agrees with a mean-field analysis as well as numerical simulations from Ref.~\cite{nahum2021}.

We also observe that if we only include noiseless simulation data from $L \leq 40$, both the parameters here $(p_c, \nu) \approx (0.33, 2.50)$, and the best fits obtained from experimental data $(p_c, \nu) \approx (0.26, 1.90)$ (the fitting procedures are described in Sec.~\ref{si:collapse}, and the results are shown in Fig.~\ref{fig:all-to-all-cross-entropy}), will result in high quality data collapses (data not shown).
This is consistent with our observation of a large uncertainty in the fitting parameters in our experimental data from Fig.~\ref{fig:all-to-all-cross-entropy}.
Indeed, collapsing the experimental data from Fig.~\ref{fig:all-to-all-cross-entropy} with the theoretical value $\nu = 2.5$, we find reasonable agreement (see Fig.~\ref{fig:all-to-all-nu_theory}), even though $\nu = 2.5$ lies outside the 90\% confidence interval, $\nu = 1.9 \pm 0.4$, as obtained from fitting procedures in Sec.~\ref{si:collapse}.

On the other hand, from our noisy data at noise rate $q = 0.1\%$, we obtain $p_c = 0.20 \pm 0.02$ and $\nu = 0.80 \pm 0.3$, see Fig.~\ref{fig:all-to-all-q-0.001-L-small}.
Recall that the same noise model and noise rate produced Fig.~\ref{fig:1D-simulated-noisy}, which are comparable to experimental results in 1D.
This observation suggests that noise affects the fitting of critical exponents more strongly in all-to-all connectivity.
We note that the qualitative trend of a decreasing $\chi$ with increasing $L$ and/or $p$ is still captured by the simple noise model at $q = 0.1\%$.


Finally, we take $\rho = \sigma$ in the noisy simulation in all-to-all connectivity, with results plotted in Fig.~\ref{fig:all-to-all-q-0.001-L-small}(c).
We find reasonable agreement with Eq.~\eqref{eq:noisy_rho=sigma_functional_form}, suggesting that the functional form may be widely applicable in a variety of different circuit connectivities.


\subsection{Statistical mechanics picture}

The qualitative behavior the results in Fig.~\ref{fig:1D-simulated-noisy} can be understood from a mapping to statistical mechanics models, which we briefly describe here. (We refer the reader to Ref.~\cite{li2023} and references therein for further details.)
Recall that
\begin{align}
    \chi \coloneqq \mathbb{E}_C \chi_C 
    = \mathbb{E}_C \frac{\sum_{\bs{m}} p_{\bs{m}}^\rho p_{\bs{m}}^\sigma}
    {\sum_{\bs{m}} \(p_{\bs{m}}^\sigma\)^2} 
    = \mathbb{E}_C \frac{\sum_{\bs{m}} \tr [C_{\bs{m}} (\rho)] \cdot
    \tr [C_{\bs{m}} (\sigma)]
    }
    {\sum_{\bs{m}} 
    (\tr [C_{\bs{m}} (\sigma)])^2
    } 
    = \mathbb{E}_C
    \frac{
        \sum_{\bs{m}} \tr [ C_{\bs{m}}^{\otimes 2} (\rho \otimes \sigma) ]
    }
    {
        \sum_{\bs{m}} \tr [ C_{\bs{m}}^{\otimes 2} (\sigma \otimes \sigma) ]
    }.
\end{align}
Here $C_{\bf m}(\rho)$ denotes the resultant state when unitaries and projective measurements (labeled by the measurement record ${\bf m}$) 
from $C$ are applied to the initial state $\rho$.\footnote{This notation is different from Ref.~\cite{li2023} to accommodate possible appearances of quantum channels.}
It is easier to study the following proxy quantity, which is an approximation of $\chi$ by averaging the numerator and the denominator separately over $C$,
\begin{align}
    \overline{\chi} =
    \frac{
        \mathbb{E}_C
        \sum_{\bs{m}} \tr [ C_{\bs{m}}^{\otimes 2} (\rho \otimes \sigma) ]
    }
    {
        \mathbb{E}_C
        \sum_{\bs{m}} \tr [ C_{\bs{m}}^{\otimes 2} (\sigma \otimes \sigma) ]
    }.
\end{align}
For $C$ a brickwork circuit with local 2-qubit random unitary gates forming a 2-design, the averages can be performed.
As a result, the numerator and the denominator will both take the form of a partition function of the Ising model on a triangular lattice, where the Boltzmann weights can be explicitly written down (\cite{andreas2019hybrid, choi2019spin, nahum2018operator, zhou2018emergent, zhou2019membrane}).
The two partition functions are identical in the bulk, and only differ in their boundary conditions (coming from the difference in initial states).
Following Ref.~\cite{li2023}, we denote them $Z_{\rho \neq \sigma}$ and $Z_{\rho = \sigma}$, respectively.

\begin{figure}
    \centering
    \includegraphics[width=0.7\linewidth]{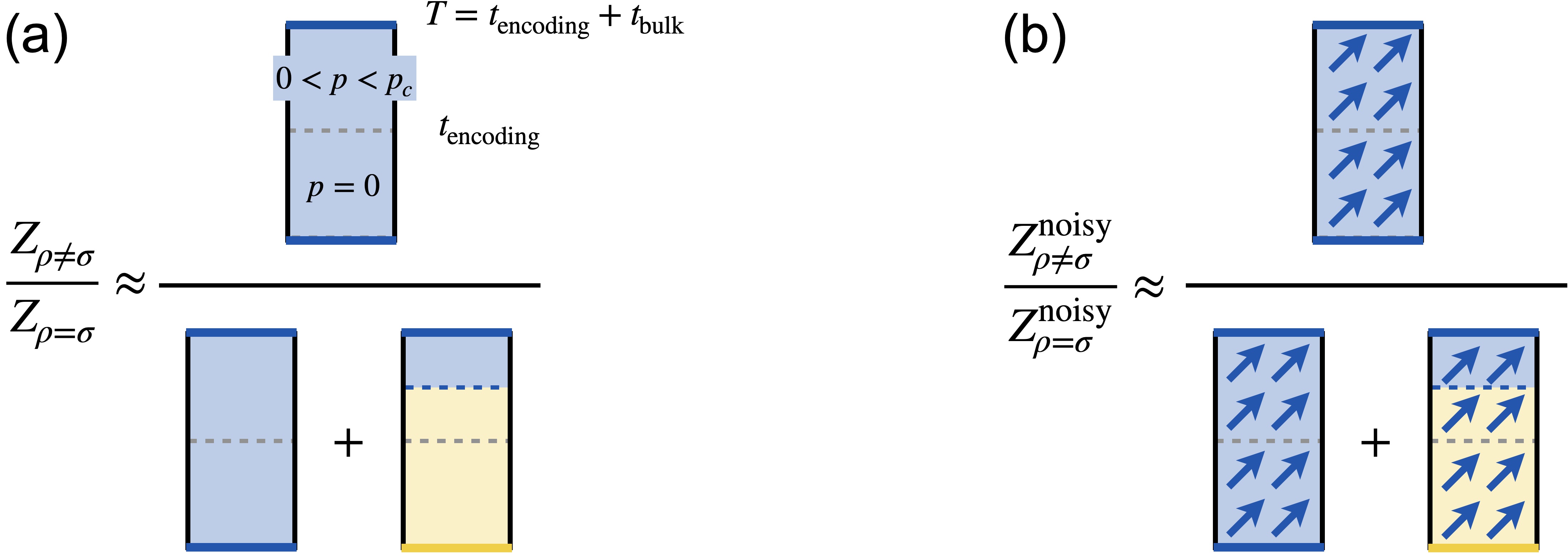}
    \caption{
    Mapping $\overline{\chi}$ defined in Eq.~\eqref{eq:chi_proxy} to quantities in an effective Ising model, when the circuit is (a) noiseless and (b) noisy.
    See the text for more details.
    In both figures the blue color represents spins pointing in the ``$+$'' direction, the yellow color represents spins pointing in the ``$-$'' direction, and the black color represents a ``free'' boundary condition, where the spins can point in either direction.}
    \label{fig:Ising-picture}
\end{figure}

In all our circuits we choose $\rho$ and $\sigma$ to be tensor products of onsite density matrices, and let them be different states.
We also take the circuit to have a purely-unitary ``encoding'' stage without measurements, before measurements take place (see Fig.~\ref{fig:circuit} of the main text).
Within these circuits, $\overline{\chi} = Z_{\rho \neq \sigma} / Z_{\rho = \sigma}$ corresponds to the partition function ratio shown in Fig.~\ref{fig:Ising-picture}(a).
Each term lives in a rectangular geometry, with the lower half an Ising model at zero temperature (corresponding to the encoding stage), and the upper half at finite temperature~\cite{li2023}.
The blue color denotes a ``$+$'' boundary condition, and the yellow color denotes a ``$-$'' one.
The numerator $Z_{\rho \neq \sigma}$ has a boundary condition where both the top and bottom spins are fixed to be $+$, whereas $Z_{\rho = \sigma}$ has an additional contribution where the bottom boundary condition is also ``$-$''.
Thus,
\begin{align}
    \label{eq:chi_proxy}
    \overline{\chi}_{\rho \neq \sigma} = \frac{Z_{\rho \neq \sigma}}{Z_{\rho = \sigma}} = \frac{1}{1+Z_{+-}/Z_{++}}.
\end{align}
The $p < p_c$ phase of circuit maps to the the ferromagnetic phase of the Ising magnet, where $- \ln (Z_{+-}/Z_{++})$ is the free energy of a horizontal domain wall separating the bottom and the top (see Fig.~\ref{fig:Ising-picture}(a)), which diverges with $L$, therefore $\overline{\chi} \to 1$.
On the other hand, in the $p > p_c$ ``paramagnetic'' phase the domain wall free energy vanishes, so $Z_{+-}/Z_{++} \to 1$ and $\overline{\chi} \to 1/2$.
We see that the numerical value of $\overline{\chi}$ in the $p>p_c$ phase differs from our numerical results, due to the annealed average.

The Ising picture is also useful for a qualitative understanding of the behavior of linear cross entropy in the presense of noise.
For simplicity, we take the the noise to be a random erasure at each spacetime location.
The cross entropy now reads
\begin{align}
    \chi \coloneqq \mathbb{E}_{C, \mathcal{N}} \chi_{C, \mathcal{N}}
    = \mathbb{E}_{C, \mathcal{N}}
    \frac{
        \sum_{\bs{m}} \tr [ (C'_{\bs{m}} \otimes C_{\bs{m}}) (\rho \otimes \sigma) ]
    }
    {
        \sum_{\bs{m}} \tr [ C_{\bs{m}}^{\otimes 2} (\sigma \otimes \sigma) ]
    }.
\end{align}
Here the circuit $C'$ is obtained from $C$ by inserting erasure noise (denoted $\mathcal{N}$) at random spacetime locations, which in general turns pure states into mixed states.
Similarly, we define
\begin{align}
    \overline{\chi}^{\rm noisy}_{\rho \neq \sigma} = 
    \frac{
    \mathbb{E}_{C, \mathcal{N}}
        \sum_{\bs{m}} \tr [ (C'_{\bs{m}} \otimes C_{\bs{m}}) (\rho \otimes \sigma) ]
    }
    {
    \mathbb{E}_{C}
        \sum_{\bs{m}} \tr [ C_{\bs{m}}^{\otimes 2} (\sigma \otimes \sigma) ]
    }
    = \frac{Z^{\rm noisy}_{\rho \neq \sigma}}{Z_{\rho = \sigma}}.
\end{align}
This quantity is similar to our of experimental data in Fig.~\ref{fig:1D-chain-cross-entropy}.
We can also consider the following ratio
\begin{align}
\label{eq:chi_rho=sigma_noisy}
    \overline{\chi}^{\rm noisy}_{\rho = \sigma} = \frac{Z^{\rm noisy}_{\rho = \sigma}}{Z_{\rho = \sigma}},
\end{align}
which approaches $1$ as the noise rate vanishes, and is similar to Fig.~\ref{fig:rho=sigma-cross-entropy}.
Both $Z^{\rm noisy}_{\rho \neq \sigma}$ and $Z^{\rm noisy}_{\rho = \sigma}$ can be obtained from their noiseless versions by applying a ``magnetic field'' everywhere in the system favoring the ``$+$'' direction and penalizing the ``$-$'' direction.
More precisely, its effect can be captured by an additional term to the energy function of the Ising model,
\begin{align}
    E[\{s\}, h] = E[\{s\}, h=0] + h \sum_j \delta_{s_j, -1} = E[\{s\}, h=0] + \frac{h}{2} \sum_j (1-s_j),
\end{align}
where $Z[h] = \tr_{\{s\}} e^{-\beta E[\{s\}, h]}$ is the Ising partition function, and $h$ is the strength of the field (proportional to the strength of the noise).
The field breaks the Ising symmetry and destroys the phase transition.
For Eq.~\eqref{eq:chi_rho=sigma_noisy}, we write the partition function in the numerator as follows,
\begin{align}
    Z^{\rm noisy}_{\rho = \sigma} = e^{- V f(h)},
\end{align}
where $V \propto L^2$ is the circuit volume, and $f(h)$ is the free energy density when the field is applied to the magnet.
Regardless of the phase the Ising magnet is in, a finite magnetization density $m(h)$ will appear, as a response to a small but finite $h$.
The free energy density can then be approximated as 
\begin{align}
    f(h) = f(h=0) + \frac{h}{2}(1-m(h)).
\end{align}
Therefore, we have
\begin{align}
    \label{eq:chi_rho=sigma_noisy_Ising}
    \overline{\chi}^{\rm noisy}_{\rho = \sigma} = \frac{Z^{\rm noisy}_{\rho = \sigma}}{Z_{\rho = \sigma}} \propto \exp\left[- {\rm const} \cdot \frac{h}{2}(1-m(h)) \cdot L^2 \right].
\end{align}
This is consistent with the functional form in Eq.~\eqref{eq:noisy_rho=sigma_functional_form} and numerical results Fig.~\ref{fig:1D-simulation-noisy-rho=sigma}(b).
In particular, we observe in Fig.~\ref{fig:1D-simulation-noisy-rho=sigma}(b) a increasing rate of the exponential decay for a fixed noise rate and increasing $p$, corresponding to a smaller magnetization $m(h)$ as we raise the temperature.

A similar exponential dependence on $L^2$ is expected for $\overline{\chi}^{\rm noisy}_{\rho \neq \sigma}$.
The dependences will cancel if we take their ratio.
As we illustrate in Fig.~\ref{fig:Ising-picture}(b), their ratio should always be upper bounded by 1,
\begin{align}
\label{eq:chi_bound_Ising}
    \frac{\overline{\chi}^{\rm noisy}_{\rho \neq \sigma}}{\overline{\chi}^{\rm noisy}_{\rho = \sigma}} = \frac{Z^{\rm noisy}_{\rho \neq \sigma}}{Z^{\rm noisy}_{\rho = \sigma}} = \frac{1}{1+ Z_{+-}(h > 0) / Z_{++}(h > 0)} \leq 1,
\end{align}
as the Ising partition functions remain  positive under the erasure channel.
To complement the statistical mechanics approach,
below in Sec.~\ref{sec:noise} we give a rigorous derivation of an upper bound of $\chi$ (rather than its proxy $\overline{\chi}$) in  circuits with stabilizer operations.
We also discuss the the apparent violation of the upper bound by experimental data.

\subsection{Effect of Pauli noise \label{sec:noise}}

Our statistical mechanics picture above suggests 
that $\chi_{\rho \neq \sigma}^{\rm noisy} \leq \chi_{\rho = \sigma}^{\rm noisy}$ (see Eq.~\eqref{eq:chi_bound_Ising}), while our experimental results clearly violate this relation, compare $\chi_{\rho = \sigma}$ (Fig.~\ref{fig:rho=sigma-cross-entropy}) with $\chi_{\rho \neq \sigma}$ (Fig.~\ref{fig:1D-chain-cross-entropy}).
To get a better handle on this, 
we formalize the following characterization of linear cross entropy in stabilizer circuits (where the circuit architecture is arbitrary).

\textbf{Proposition 1.} Let $\rho$ and $\sigma$ be two stabilizer states, which are in general different from each other.
Consider a ``noiseless'' Clifford circuit $C$, composed of arbitrary Clifford unitaries and arbitrary Pauli measurements; and a noisy Clifford circuit $C'$ obtained from $C$ by injecting a number of stabilizer channels\footnote{With stabilizer channels we mean channels that can be represented as stabilizer operations, of the form $\mathcal{E}(\cdot) = \frac{1}{2}(\cdot) + \frac{1}{2} P (\cdot) P$ with Pauli operator $P$ (e.g. biased erasure errors), or their compositions (e.g. erasure errors).} at arbitrary spacetime locations.
Define
\begin{align}
\label{eq:linear_XEB_def}
    \chi(C', \rho | C, \sigma)
    =
    \frac{
        \sum_{\mathbf{m}} \mathrm{tr} [C'_{\bf m}(\rho)] \cdot  \mathrm{tr} [C_{\bf m}(\sigma)].
    }
    {
        \sum_{\mathbf{m}} (\mathrm{tr} [C_{\bf m}(\sigma)])^2
    }.
\end{align}
We have the following inequality between the linear cross entropies:
\begin{align}
\label{eq:inequality_S29}
    \chi(C', \rho | C, \sigma) \leq \chi(C', \sigma | C, \sigma).
\end{align}

\textit{Proof.}
We first adopt a purified representation for the Pauli measurements, see Appendix S2 of~\cite{li2023}.
For each measurement of Pauli operator $P$ in the circuit, we can introduce an additional register qubit, and apply a controlled Clifford unitary operator acting on the register qubit as well as qubits being measured, followed by a dephasing channel on the register, to simulate the effect of that measurement.
In effect, at the end of the time evolution we have the following joint stabilizer states on the physical qubits $Q$ and the register qubits $R$,
\begin{align}
    \rho^{C'}_{QR} = \sum_{\bs{m}} C'_{\bs{m}}(\rho) \otimes \ket{\bs{m}}\bra{\bs{m}}_R, \\
    \sigma^{C'}_{QR} = \sum_{\bs{m}} C'_{\bs{m}}(\sigma) \otimes \ket{\bs{m}}\bra{\bs{m}}_R, \\
    \sigma^{C}_{QR} = \sum_{\bs{m}} C_{\bs{m}}(\sigma) \otimes \ket{\bs{m}}\bra{\bs{m}}_R.
\end{align}
With this representation, we have
\begin{align}
    \chi(C', \sigma | C, \sigma) = \frac{{\rm tr} [\sigma^{C'}_R \cdot \sigma^{C}_R]}{{\rm tr} [(\sigma^{C}_R)^2]}, \\
    \chi(C', \rho | C, \sigma) = \frac{{\rm tr} [\rho^{C'}_R \cdot \sigma^{C}_R]}{{\rm tr} [(\sigma^{C}_R)^2]},
\end{align}
where $\rho^{C'}_R$, $\sigma^{C'}_R$, and $\sigma^{C}_R$ are reduced state of $\rho^{C'}_{QR}$, $\sigma^{C'}_{QR}$, and $\sigma^{C}_{QR}$ on $R$, respectively.

Denote by $\mathcal{S}(\rho)$ the stabilizer group corresponding to a stabilizer state $\rho$.
By induction, one can show that (Lemma 1, see below)
\begin{align}
\label{eq:stabilizer_subgroup_relation}
    \mathcal{S}({\sigma^{C'}_{QR}}) \subset \mathcal{S}({\sigma^{C}_{QR}}),
\end{align}
due to that $C'$ is obtained from $C$ by additional stabilizer channels.
Such channels can only eliminate elements from the stabilizer group.

By similar reasoning, we also have (Lemma 2, see below)
\begin{align}
    \mathcal{S}({\rho^{C'}_{QR}}) \cap \mathcal{S}({\sigma^{C}_{QR}}) \subset 
    \mathcal{S}({\sigma^{C'}_{QR}}) \cap \mathcal{S}({\sigma^{C}_{QR}}) = \mathcal{S}({\sigma^{C'}_{QR}})
\end{align}
We can then calculate
\begin{align}
    \chi(C', \rho | C, \sigma) = \frac{{\rm tr} [\rho^{C'}_R \cdot \sigma^{C}_R]}{{\rm tr} [(\sigma^{C}_R)^2]} 
    \leq \frac{|\mathcal{S}({\rho^{C'}_{QR}}) \cap \mathcal{S}({\sigma^{C}_{QR}})|}{|\mathcal{S}({\sigma^{C}_{QR}})|}
    \leq \frac{|\mathcal{S}({\sigma^{C'}_{QR}}) \cap \mathcal{S}({\sigma^{C}_{QR}})|}{|\mathcal{S}({\sigma^{C}_{QR}})|} = \chi(C', \sigma | C, \sigma)
\end{align}
where we use the following result~\cite{li2023}
\begin{align}
    \tr [\rho_R \sigma_R]
     =&\  \frac{1}{2^{2|R|}} \sum_{g \in \mathcal{S}_{\rho_R}} \sum_{h \in \mathcal{S}_{\sigma_R}} \tr [gh]
    = 
    \begin{cases}
        2^{-|R|} |\mathcal{S}_{\rho_R} \cap \mathcal{S}_{\sigma_R} |, & -1 \notin \mathcal{S}_{\rho_R} \cdot \mathcal{S}_{\sigma_R} \\
        0, & -1 \in \mathcal{S}_{\rho_R} \cdot \mathcal{S}_{\sigma_R}
    \end{cases}
\end{align}
for any two stabilizer states on $R$.
\hfill $\Box$

~

\textbf{Lemma 1.} $\mathcal{S}({\sigma^{C'}_{QR}}) \subset \mathcal{S}({\sigma^{C}_{QR}})$.

\textit{Proof.} 
We can show this by induction on the quantum operations appearing in the circuit $C'$ and $C$, which are Clifford unitaries and stabilizer channels.
For brevity, we adopt the shorthand notation $\mathcal{S}'_t$ and $\mathcal{S}_t$ to denote the stabilizer groups of the instantaneous states for the two circuits $C'$ and $C$ after $t$ quantum operations shared between $C$ and $C'$ are applied.
\begin{itemize}
    \item At initialization, before any operation is applied, the two states are equal, so that $\mathcal{S}_0' = \mathcal{S}_0 \subset \mathcal{S}_0$.
    \item If the next operation is a unitary shared between $C$ and $C'$, we have that
    \begin{align}
        \mathcal{S}'_t \to \mathcal{S}'_{t+1} = U \mathcal{S}'_t U^\dagger, \quad \mathcal{S}_t \to \mathcal{S}_{t+1} = U \mathcal{S}_{t+1} U^\dagger.
    \end{align}
    The inclusion is prevserved.
    \item If the next operation is a stabilizer channel shared between $C$ and $C'$ we have
    \begin{align}
    \label{eq:stabilizer_channel}
        \mathcal{E}(\cdot) = \frac{1}{2}(\cdot) + \frac{1}{2} P (\cdot) P,
    \end{align}
    then
    \begin{align}
        \mathcal{S}'_t \to&\ \mathcal{S}'_{t+1} = \{g \in \mathcal{S}' : gP = Pg\} \subset \mathcal{S}'_t, \\
        \mathcal{S}_t \to&\ \mathcal{S}_{t+1} = \{g \in \mathcal{S} : gP = Pg\} \subset \mathcal{S}_t.
    \end{align}
     The inclusion is also prevserved.
     \item If the next operation is a stabilizer channel that is only in $C'$ but not in $C$, $\mathcal{S}$ remains unchanged, and we have
     \begin{align}
          \mathcal{S}'_t \to&\ \{g \in \mathcal{S}'_t : gP = Pg\} \subset \mathcal{S}'_t \subset \mathcal{S}_t.
     \end{align}
\end{itemize}
By induction, we conclude that $\mathcal{S}({\sigma^{C'}_{QR}}) \subset \mathcal{S}({\sigma^{C}_{QR}})$ after all operations are applied. 
\hfill $\Box$

~

\textbf{Lemma 2.} $\mathcal{S}({\rho^{C'}_{QR}}) \cap \mathcal{S}({\sigma^{C}_{QR}}) \subset 
    \mathcal{S}({\sigma^{C'}_{QR}}) \cap \mathcal{S}({\sigma^{C}_{QR}})$.

\textit{Proof.}
The idea is similar to the proof of Lemma 1.
Let the corresponding stabilizer groups at time $t$ be denoted $\mathcal{S}'_\rho$, $\mathcal{S}'_\sigma$, and $\mathcal{S}_\sigma$, respectively.
We want to show that $\mathcal{S}'_\rho \cap \mathcal{S}_\sigma \subset \mathcal{S}'_\sigma \cap \mathcal{S}_\sigma$ at all times, by induction.
This property is true at initialization since $\mathcal{S}'_\sigma  = \mathcal{S}_\sigma$ to start with.
The preservation of this property under Clifford unitaries and quantum operations can also be straightforwardly verified.
\hfill $\Box$

~

The inequality Eq.~\eqref{eq:inequality_S29}  applies to each sample from the ensemble of stabilizer circuits we considered in our numerical simulation above.
Eq.~\eqref{eq:inequality_S29} also applies to a slighlty broader class of error channels beyond stabilizers, e.g. if each additional stabilizer channel in $C'$ (but not in $C$) is replaced by a probabilistic mixture of stabilizer channels, since linear cross entropy is linear in each channel, see Eq.~\eqref{eq:linear_XEB_def}.
These include weak depolarizing or weak dephasing noise, as is usually assumed in the literature of random circuit sampling~\cite{google2019supremacy, Bouland_2022, ware2023sharp, morvan2023phase}.

Furthermore, we argue that the same inequality holds for a compressed circuit, as obtained from the algorithm in Sec.~\ref{si:compression}.
Recall that the compression algorithm returns a circuit composed of measurements of Pauli operators only.
In particular, a Pauli measurement in the uncompressed circuit either maps to another Pauli measurement in the compressed circuit, or to a classical coin flip.
In the compressed circuit, the sign of a later Pauli operators might depend on earlier measurement results and/or earlier coin flip results, therefore ``adaptive''.
To remove such adaptivity, we showed that one can simply assume that all Pauli operators (to be measured) have the sign $+1$, and the adaptivity can be equivalently achieved by postprocessing the measurement results and the results of the classical coin flips.
In particular, the postprocessing takes the form of an $\mathbb{F}_2$-linear map on $(\mathbb{F}_2)^{N}$, where $N$ is the number measurements in the uncompressed circuit (which is equal to the number of measurements plus the number of coin flips in the compressed circuit).
Thus the compressed circuit would seem to have more structure than the circuits $C$ we considered above.
To make our results applicable, we note that
\begin{itemize}
    \item The classical coin flips can be simulated by initializing a corresponding register qubit in the maximally mixed state.
    We treat the other Pauli measurements as usual, by introducing an register qubit for each.
    This way, we have $N$ register qubits in total.
    \item The ``postprocessing'' map can be realized by a Clifford unitary on the $N$ register qubits, which is also diagonal in the computational basis.
\end{itemize}
This way, the compressed circuits can also be recasted as a stabilizer circuit with Clifford unitaries and stabilizer channels.
For such a circuit $C$, and a ``noisy'' version $C'$ obtained from $C$ by inserting stabilizer channels (which can be seen as a crude approximation of the hardware experiments we carried out), we will also have
\begin{align}
    \chi(C', \rho | C, \sigma) \leq \chi(C', \sigma | C, \sigma).
\end{align}

The differences between the setup in our Proposition and the experiment are (i) the initial state $\rho$ in our experiments is taken to be a nonstabilizer state, and (ii) the noise in our experiments is not simply of the form of a stabilizer channel (or their probabilistic mixture).
The violation of the inequality Eq.~\eqref{eq:inequality_S29} by our experimental results in Figs.~\ref{fig:rho=sigma-cross-entropy},\ref{fig:1D-chain-cross-entropy} may thus be attributed to non-stabilizerness of the initial state, realistic error models (which necessarily involve coherent and non-unital noise, as well as read-out error), or a combination thereof.
While it is easy to construct contrived example Clifford circuits with adversarial coherent noise that show violations of the bound (e.g. a unitary that exchanges the two initial state), we have not been able to find natural and physically relevant examples that can closely approximate the experimental data.
It will be an interesting future direction to explore the effects and the description of non-stabilizer noise channels on MIPT, and conversely, the extent to which many-body phenomena in random circuits can be informative of noise.

~

We also state a result similar in spirit to Proposition 1, which applies to other choices of initial states.
The key condition to Proposition 1 is the relation $\mathcal{S}({\rho^{C'}_{QR}}) \cap \mathcal{S}({\sigma^{C}_{QR}}) \subset \mathcal{S}({\sigma^{C'}_{QR}}) \cap \mathcal{S}({\sigma^{C}_{QR}})$, which is preserved throughout the time evolution.
It is straightforward to a similar condition in the following scenario.

\textbf{Proposition 2.} Let $\rho_{1,2}$ and $\sigma$ be three stabilizer states, which are in general different from each other.
Let $\rho_1$ be obtainable from $\rho_2$ via stabilizer channels (or their probabilistic mixtures).
Consider a ``noiseless'' Clifford circuit $C$ and a noisy Clifford circuit $C'$ specified the same way as in Proposition 1.
We have the following inequality between the linear cross entropies:
\begin{align}
\label{eq:inequality_S45}
    \chi(C', \rho_1 | C, \sigma) \leq \chi(C', \rho_2 | C, \sigma).
\end{align}

\section{Qubit selection}\label{si:qubit_selection}

For the 1D-chain experiment with $\rho \neq \sigma$, the qubits were selected heuristically at run time as in Ref. \cite{koh2022}. The qubits we selected based on the one and two qubit gate error rates, $\epsilon^{1\mathrm{q}}$ and $\epsilon^{2\mathrm{q}}$, respectively, as well as the qubit readout error $\epsilon^{\mathrm{ro}}$ rates provided by IBM in their hardware callibration data. Denoting by $\chi$ the set of qubit selections which contain all $L/2$ qubits in a connected chain, an average circuit error for circuit $C$ is calculated as 
\begin{equation}
    \mathcal{E}_{x\in\chi}[C]=\sum_{j\in x}(\epsilon^{1\mathrm{q}}N_j^{1\mathrm{q}}[C]+\epsilon^{2\mathrm{q}}N_j^{2\mathrm{q}}[C]+\epsilon^{\mathrm{ro}}N_j^{\mathrm{ro}}[C]),
\end{equation}
where the subscript $j$ represents the $j$'th qubit in the qubit set $x$, and the function $N_j^{\mathrm{1q,2q,ro}}$ computes the number of single qubit gates, two qubit gates, and measurements, respectively, acting on qubit $j$ in the circuit $C$. The qubit chain used in the experiment is then selected as the one which minimizes the average error over all circuits $\mathcal{C}$, $\mathrm{argmin}_{x\in\chi}\expect_{C\in \mathcal{C}}{\mathcal{E}_x[C]}$. 
For the all-to-all and $\rho=\sigma$ experiments, we used the same qubit layouts that were selected for the 1D-chain.

In Table~\ref{table:device} we list the error rates on single-qubit measurements, 1-qubit gates, and 2-qubit gates for the device used in this study, $\emph{ibm}\_\emph{sherbrooke}$. We note that interpreting these data in terms of the average depolarizing strength $q$ in a circuit requires modeling which is beyond the scope of this work.

The single-qubit readout error is estimated as 
\begin{equation}
\epsilon^{\mathrm{ro}} = \frac{\mathrm{P(meas=0|prep=1)}+\mathrm{P(meas=1|prep=0)}}{2}
\end{equation}
i.e., as the average of the probabilities of measuring a qubit in the state $|x\rangle$ (with $x=0,1$) conditional to its preparation in $|(x + 1) \% 2 \rangle$, where $\% 2$ denotes sum modulo $2$ (i.e. for $x=0,1$ $(x+1)\% 2=1,0$). 

The 1-qubit gates available are 
\begin{equation}
I = \begin{pmatrix} 1 & 0 \\ 0 & 1 \end{pmatrix}
\;,\;
X = \begin{pmatrix} 0 & 1 \\ 1 & 0 \end{pmatrix}
\;,\;
\sqrt{X} = \frac{1}{2} \begin{pmatrix} 1+i & 1-i \\ 1-i & 1+i \end{pmatrix}
\;.
\end{equation}
All other single-qubit gates can be constructed by combining these operations with Pauli $Z$ rotations, which are implemented virtually on superconducting quantum devices~\cite{mckay2017efficient}, and thus are error-free.
The 2-qubit gate available is the echoed cross-resonance gate~\cite{sheldon2016procedure},
\begin{equation}
ECR = \frac{1}{\sqrt{2}}
\begin{pmatrix}
 0 & 1 &  0 & i \\
 1 & 0 & -i & 0 \\
 0 & i &  0 & 1 \\
-i & 0 &  1 & 0 \\
\end{pmatrix}\;,
\end{equation}
The ECR gate is maximally entangling and is equivalent to a CNOT gate up to single-qubit rotations. Errors that affect the 1- and 2-qubit gates are estimated using the randomized benchmarking procedure~\cite{magesan2011scalable}. The median 2-qubit error rate in Table~\ref{table:device} is in line with published results for Eagle R3 devices~\cite{mckay2023benchmarking}.

\begin{table}
\begin{tabular}{cccccc}    
\hline\hline
operation & mean error & standard deviation & median error & min error & max error \\
\hline
Readout ($\epsilon^{\mathrm{ro}}$)            & 0.0330 & 0.0500 & 0.0158 & 0.0024 & 0.3507 \\
P(meas=0$|$prep=1)                            & 0.0425 & 0.0866 & 0.0184 & 0.0038 & 0.6996 \\
P(meas=1$|$prep=0)                            & 0.0235 & 0.0454 & 0.0092 & 0.0010 & 0.4244 \\
\hline
$\mathsf{I}$                                  & 0.0005 & 0.0013 & 0.0002 & 0.0001 & 0.0135 \\
Pauli $X$ ($\epsilon^{\mathrm{1q}}$) & 0.0005 & 0.0013 & 0.0002 & 0.0001 & 0.0135 \\
$\sqrt{X}$                           & 0.0005 & 0.0013 & 0.0002 & 0.0001 & 0.0135 \\
\hline
$ECR$  ($\epsilon^{\mathrm{2q}}$)    & 0.0121 & 0.0147 & 0.0077 & 0.0027 & 0.1148 \\
\hline\hline
\end{tabular}
\caption{Mean error (column 2), standard deviation on the mean error (column 3), median error (column 4), minimum error (column 5), and maximum error (column 6) for the quantum operations available on the $\emph{ibm}\_\emph{sherbrooke}$ device. We list errors on single-qubit measurements (readout, and probability to measure a qubit in the state $|x\rangle$ with $x=0,1$ conditional to its preparation in $|(x + 1) \% 2 \rangle$) in lines 1-3; errors on 1-qubit gates (identity, Pauli $X$, and square root thereof) in lines 4-6; and errors on the native 2-qubit gate ($ECR$ for echoed cross-resonance gate) in line 7.} \label{table:device}
\end{table}

\section{Fitting parameters \texorpdfstring{$\nu$}{nu} and \texorpdfstring{$p_c$}{pc} by collapsing hardware data} \label{si:collapse}
Near the critical measurement rate $p_c$, the order parameter $\chi$ for different system sizes and under suitable rescaling is expected to collapse onto a single curve \cite{nahum2018hybrid, stanley1999, bhattacharjee2001}. Quantitatively, this can be expressed as $\chi(L, p)$ collapsing to the same curve for all system sizes $L$ when we suitably rescale both $L$ and $p$:
\begin{equation}
    \chi(L,p)=F\left[L^{1/\nu}(p-p_c)\right].
\end{equation}
The critical measurement rate depends on the microscopic details of the circuits, such as the encoding and bulk ratios, whereas the the critical exponent is independent of the microscopic circuit details and is the same for all systems in the same universality class \cite{nahum2018hybrid, stanley1999}. If the scaling function $F$ was known, we could obtain the optimal $p_c$ and $\nu$, denoted by $p_c^*$ and $\nu^*$, by minimizing the residual sum of squares (RSS) over all data points:
\begin{equation}
    p_c^*, \nu^* = \underset{p_c,\nu}{\mathrm{arg\,min}}\sum_L\sum_p\left(F\left[L^{1/\nu}(p-p_c)\right] - \chi_{\mathrm{exp}}(L,p)\right)^2,
\end{equation}
where $\chi_{\mathrm{exp}}(L,p)$ is the cross entropy obtained from the experiment for a system size $L$ and measurement rate $p$. When the scaling function is unknown, we still find $p_c^*$ and $\nu^*$ by minimizing an RSS, but instead use an interpolating function for our scaling function for a fixed $L$, followed by symmetrization over all $L$ in order to prevent preferential treatment of any portion of the data \cite{bhattacharjee2001, koh2022}. Our approach to fitting $p_c$ and $\nu$ follows Ref. \cite{koh2022} with modifications due to there being only one critical exponent in our case, versus two critical exponents in Ref. \cite{koh2022}. We denote by $\mathcal{L}$ the set of system sizes used in the experiment and $\mathcal{P}_L$ the set of measurement rates used for a fixed $L$. For each $L\in\mathcal{L}$ and $p\in\mathcal{P}_L$, we first compute the rescaled controlled variable
\begin{equation}
    q_L(p)=L^{1/\nu}(p-p_c).
\end{equation}
We then construct an interpolating function for $\chi(L,p)$ from the rescaled experimental data, which we denote by $f_L(q)$. The interpolating function is used since the $q$ values for different values of $L$ are different, and the RSS is taken over points with identical $q$ values. From numerical simulations, we expect the scaling function to decrease monotonically for increasing $p$ \cite{li2023}. To preserve this monotonicity, we use a piecewise cubic Hermite polynomial implemented in SciPy to construct the interpolating function \cite{scipy}. We denote the set of $q_L$ as $\mathcal{Q}_L$, $q^-_L =\min_{\mathcal{Q}_L}$ and $q^+_L = \max_{\mathcal{Q}_L}$. Adapting the measure of goodness of fit from References \cite{bhattacharjee2001} and \cite{koh2022}, we define the loss function as 
\begin{equation}
    \label{eq:parameter_cost_function}
    R(\nu,p_c)=\sum_{L\in\mathcal{L}}\sum_{\substack{L'\in\mathcal{L},\\L'\neq L}}\sum_{\substack{q\in\mathcal{Q}_{L'}, \\ q^-_L\leq q \leq q^+_L}}\left(f_L(q) - f_{L'}(q)\right)^2.
\end{equation}


In the innermost summation, we constrain $q$ by  $q^-_L\leq q \leq q^+_L$ in order to avoid extrapolation of $f_L(q)$. Our reported best fit values of $p_c$ and $\nu$ are then given by \begin{equation}
    p_c^*, \nu^* = \underset{p_c,\nu}{\mathrm{arg\,min}}\ R(\nu,p_c).
\end{equation}

Following References \cite{koh2022, bhattacharjee2001},the errors for $\nu$ and $p_c$ are given by the width of the minimum at level $\eta$:
\begin{align}
    \delta\nu_{\pm}&=\eta\nu^*\left[2\log{\frac{R(\nu^*\pm\eta\nu^*,p_c^*)}{R(\nu^*,p_c^*)}}\right]^{-1/2}\\
    \delta {p_c}_{\pm}&=\eta p_c^*\left[2\log{\frac{R(\nu^*,p_c^*\pm \eta p_c^*)}{R(\nu^*,p_c^*)}}\right]^{-1/2}.
\end{align}
Our final values of $\nu$ and $p_c$ are then reported, setting $\eta$ to the 10\% level, as 
\begin{align}
    \nu^* &\pm \max(\delta\nu_+,\delta\nu_-)\\
    p_c^* &\pm \max(\delta{p_c}_+,\delta{p_c}_-)
\end{align}

\begin{figure}[ht!]
    \centering
    \includegraphics[width=\textwidth]{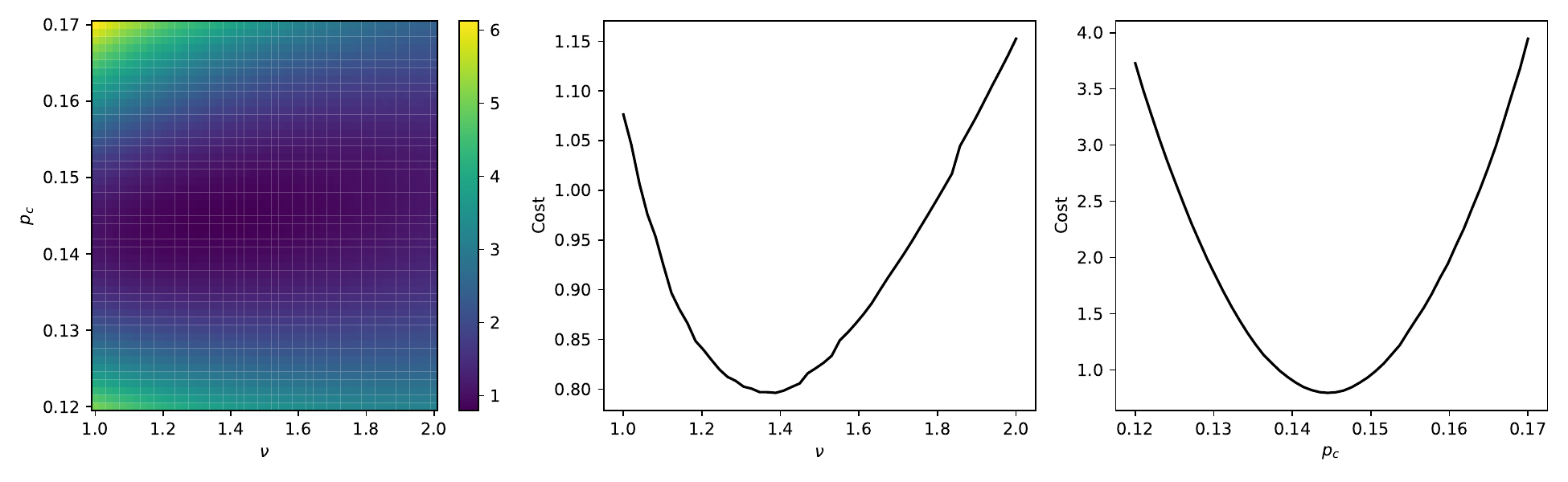}
\caption{(a) The cost function as defined in Equation~\eqref{eq:parameter_cost_function} for the 1D chain with varying $\nu$ and $p_c$. (b) The cost function when $p_c$ is held fixed at its optimal value and $\nu$ is varied. (c) The cost function when $\nu$ is held fixed at its optimal value and $p_c$ is varied.}
\label{fig:cost_function_1D}
\end{figure}

\begin{figure}[ht!]
    \centering
    \includegraphics[width=\textwidth]{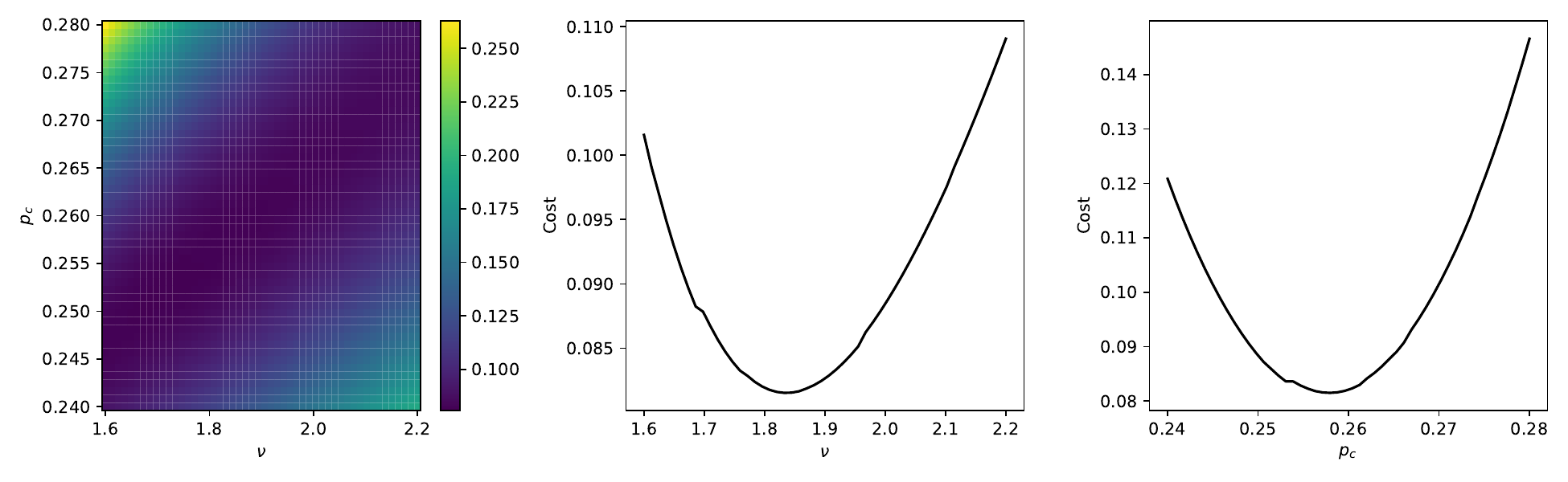}
\caption{(a) The cost function as defined in Equation~\eqref{eq:parameter_cost_function} for the all-to-all system with varying $\nu$ and $p_c$. (b) The cost function when $p_c$ is held fixed at its optimal value and $\nu$ is varied. (c) The cost function when $\nu$ is held fixed at its optimal value and $p_c$ is varied.}
\label{fig:cost_function_all-to-all}
\end{figure}

In Figure \ref{fig:cost_function_1D}(a) the cost function for the 1D chain is shown in a parameter space near the optimum. The reported uncertainties corresponding to the 90\% confidence interval are the width of the minima when we increase the cost function by 10\%. Figure \ref{fig:cost_function_1D}(b) (\ref{fig:cost_function_1D}(c)) shows the cost function for the 1D chain when $p_c$ ($\nu$) is held fixed at its minimum and $\nu$ ($p_c$) is varied. Figure \ref{fig:cost_function_all-to-all} shows the corresponding cost function for the all-to-all system. 

The low reported uncertainty in the value of $p_c$ is a consequence of the cost function having a sharp minimum in the $p_c$ direction (see Figures \ref{fig:cost_function_1D}(b) and \ref{fig:cost_function_all-to-all}(b)), and the relatively large uncertainty for $\nu$ results from the cost function having a broad minimum (see Figures \ref{fig:cost_function_1D}(c) and \ref{fig:cost_function_all-to-all}(c)).

\section{Calculation of error bars} \label{si:stats}

In order for the linear cross entropy to be a scalable probe for measurement induced phase transitions, the number of circuits and and circuit evaluations required for a given $(L,p)$ pair must be polynomial in $L,\ p$, and $1/\epsilon$, the error in estimating $\chi(L,p)$ from multiple samples.  As shown in Reference \cite{li2023}, the number of samples can in fact be taken to be independent of $L$ and $p$, and exhibits a linear dependence on $N$ in $1/\epsilon$, where $N$ is the number of circuits used. We can see this dependence explicitly in the calculation of the error bars reported in the main text, shown in the following. 

For a given $(L,p)$ pair, we use $N$ randomly generated circuits and execute each circuit $M$ times on IBM's quantum hardware, resulting in $M$ different measurement outcomes. We calculate the cross entropy for each circuit $i$ as
\begin{equation}
    \chi_i = \frac{1}{M}\sum_{j=1}^Mx_{ij},
\end{equation}
where $x_{ij}$ is the $j$'th measurement bit string for the $i$'th circuit and is defined as
\begin{equation}
    x_{ij}=\begin{cases}
        1, \text{ if } x_{ij} \text{ can occur on } \sigma_i  \\ 
        0, \text{ if } x_{ij} \text{ cannot occur on } \sigma_i
    \end{cases}.
\end{equation}
Here, $\sigma_i$ is the $\sigma$ circuit corresponding to the $i$'th $\rho$ circuit. We next calculate the standard error of the mean as \begin{equation}
    \epsilon_i=\frac{\hat{s}_i}{\sqrt{M}},\quad \hat{s}_i^2=\frac{1}{M-1}\sum_{j=1}^M(x_{ij} - \chi_i)^2.
\end{equation}

We then compute the final estimate of the cross entropy as $\bar{\chi}=(1/N)\sum_{i=1}^N\chi_i$. The variance of $\bar{\chi}$ is given by \begin{equation}
    \epsilon^2=\frac{1}{N}\sum_{i=1}^N\epsilon_i^2
\end{equation}
and the error bars reported in all figures are given by $\bar{\chi}\pm 1.96\epsilon$, representing the 95\% confidence interval for the estimate of $\chi$.

\section{Error mitigation for hardware experiments} \label{si:hardware}

\subsection{Dynamical decoupling}

\begin{figure}[ht!]
\includegraphics[width=\textwidth]{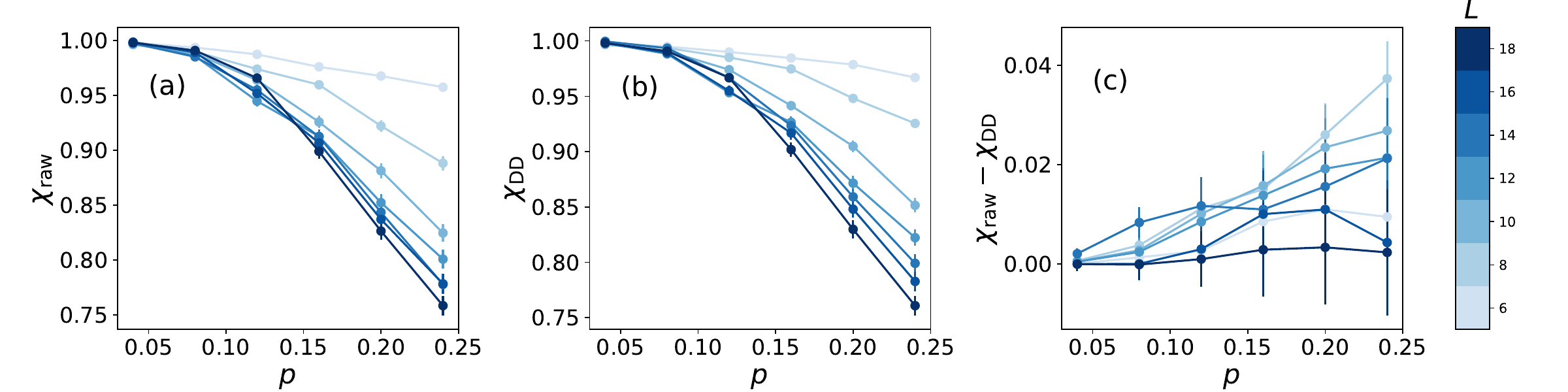}
\caption{Cross entropy $\chi$ for chains of $L=6$ to $L=18$ qubits, with initial states $\rho=\sigma$, computed without (a) and with (b) dynamical decoupling, and difference between these two quantities (c).}
\label{fig:DD}
\end{figure}

Dynamical decoupling (DD) is a quantum control technique employed in quantum computing to mitigate errors by taking advantage of time-dependent pulses \cite{viola1998dynamical,kofman2001universal,biercuk2009optimized,rost2020simulation,niu2022effects,niu2022analyzing,ezzell2022dynamical}. In its simplest form, DD is implemented by sequences of $X$ control pulses, whose effect is to protect qubits from decoherence due to low-frequency system-environment coupling. Here, we applied sequences of two $X$ pulses (as in  Ramsey echo experiments) to idle qubits.
In Figure ~\ref{fig:DD}, we illustrate the impact of DD on the cross entropy, focusing on the $\rho=\sigma$ case. As seen, for $L\simeq 10$, DD increases the cross entropy towards the exact value of $\chi=1$. However, the increase in $\chi$ is of order 0.01 whereas the difference between $\chi$ and 1 is of order 0.1 and, furthermore, it becomes less pronounced for $L \simeq 18$.

\subsection{Readout error mitigation}
\begin{figure}[htb!]
    \centering
    \includegraphics[width=\linewidth]{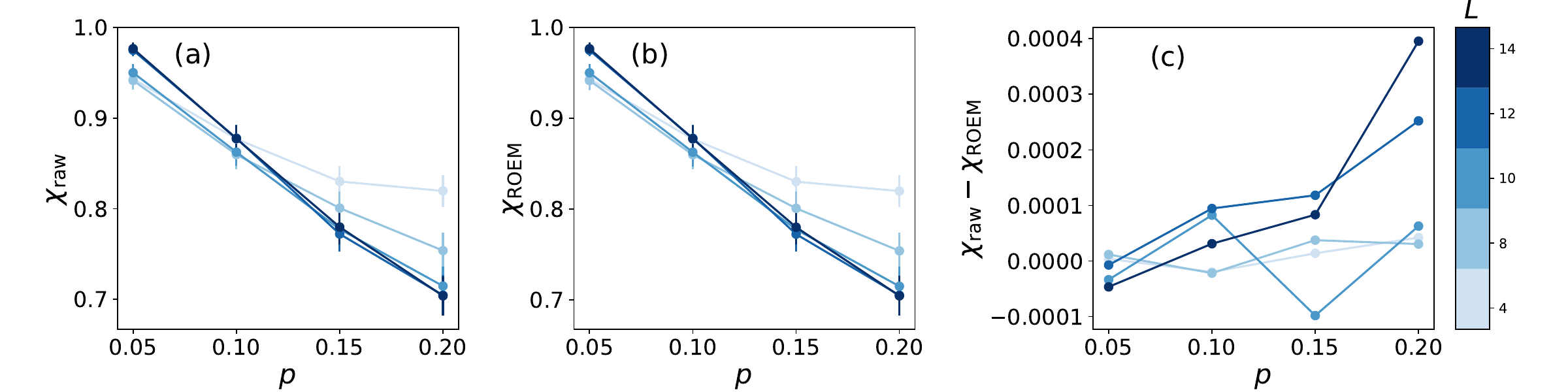}
    \caption{Effects of readout error mitigation on cross entropy for systems with up to 7 physical qubits. (a) The raw cross entropies without ROEM. (b) The cross entropies with ROEM applied. (c) The difference $\chi_{\mathrm{raw}} - \chi_{\mathrm{ROEM}}$, which shows that the differences between the raw and ROEM cross entropies are significantly smaller than the error bars for the raw cross entropies.}
    \label{fig:roem}
\end{figure}

Readout error mitigation (ROEM) is a standard technique to compensate for errors incurred during qubit readout \cite{bravyi2021, nation2021, PhysRevA.106.012423}. We tested ROEM for small systems of up to $L=14$ (7 physical qubits) and observed negligible differences between the readout error mitigated cross entropies and the unmitigated cross entropies, see Figure \ref{fig:roem}. Due to the negligible effects of ROEM, we did not use ROEM for any of the results presented in the main text. 


\end{document}